\documentclass[twocolumn,tighten,twocolappendix]{aastex701}
\usepackage{graphicx}
\usepackage{amsmath}
\usepackage{amsfonts}
\usepackage{amssymb}
\usepackage{xcolor}

\newcommand{\sunrise}{{\sc{Sunrise~iii}}}

\begin{document}

\title{Solar flare ribbons structured by uncombed chromospheric loops} 

\author[0000-0002-9270-6785]{L. P. Chitta}
\affiliation{Max-Planck-Institut für Sonnensystemforschung, Justus-von-Liebig-Weg 3, 37077 Göttingen, Germany}
\email[show]{chitta@mps.mpg.de}  

\author[0000-0003-3621-6690]{E. R. Priest}
\affiliation{School of Mathematics and Statistics, University of St Andrews, St Andrews, KY16 9SS, UK}
\email{eric.r.priest@gmail.com}


\author[orcid=0000-0001-8829-1938,sname='Orozco~Suárez']{David~Orozco~Suárez} \affiliation{Instituto de Astrofísica de Andalucía, CSIC, Glorieta de la Astronomía s/n, 18008 Granada, Spain}\affiliation{Spanish Space Solar Physics Consortium}\email{orozco@iaa.es}
\author[orcid=0000-0003-0175-6232,sname='Siu-Tapia']{Azaymi~L.~Siu-Tapia} \affiliation{Instituto de Astrofísica de Andalucía, CSIC, Glorieta de la Astronomía s/n, 18008 Granada, Spain}\affiliation{Spanish Space Solar Physics Consortium}\email{siu@iaa.es}
\author[orcid=0000-0002-3387-026X,sname='del~Toro~Iniesta']{Jose~Carlos~del~Toro~Iniesta} \affiliation{Instituto de Astrofísica de Andalucía, CSIC, Glorieta de la Astronomía s/n, 18008 Granada, Spain}\affiliation{Spanish Space Solar Physics Consortium}\email{jti@iaa.es}
\author[orcid=0000-0002-7318-3536,sname='Bailén']{Francisco~Javier~Bailén} \affiliation{Instituto de Astrofísica de Andalucía, CSIC, Glorieta de la Astronomía s/n, 18008 Granada, Spain}\affiliation{Spanish Space Solar Physics Consortium}\email{fbailen@iaa.es}
\author[orcid=0000-0002-2055-441X,sname='Blanco~Rodríguez']{Julian~Blanco~Rodríguez} \affiliation{Universitat de Valencia Catedrático José Beltrán 2, E-46980 Paterna-Valencia, Spain}\affiliation{Spanish Space Solar Physics Consortium}\email{julian.blanco@uv.es}
\author[orcid=0000-0001-9228-3412,sname='Álvarez-Herrero']{Alberto~Álvarez-Herrero} \affiliation{Instituto Nacional de T\'ecnica Aeroespacial (INTA), Ctra. de Ajalvir, km. 4, E-28850 Torrejón de Ardoz, Spain}\affiliation{Spanish Space Solar Physics Consortium}\email{alvareza@inta.es}

\author[orcid=0000-0003-4738-7727,sname='Balaguer~Jiménez']{Maria~Balaguer~Jiménez} \affiliation{Instituto de Astrofísica de Andalucía, CSIC, Glorieta de la Astronomía s/n, 18008 Granada, Spain}\affiliation{Spanish Space Solar Physics Consortium}\email{balaguer@iaa.es}
\author[orcid=0000-0002-4208-3575,sname='Sanchis~Kilders']{Esteban~Sanchis~Kilders} \affiliation{Universitat de Valencia Catedrático José Beltrán 2, E-46980 Paterna-Valencia, Spain}\affiliation{Spanish Space Solar Physics Consortium}\email{esteban.sanchis@uv.es}
\author[orcid=0000-0001-9272-6439,sname='Torralbo']{Ignacio~Torralbo} \affiliation{Universidad Politécnica de Madrid,  Plaza Cardenal Cisneros 3, E-28040 Madrid, Spain}\affiliation{Spanish Space Solar Physics Consortium}\email{ignacio.torralbo@upm.es}
\author[orcid=0000-0002-3242-1497,sname='Kuckein']{Christoph~Kuckein} \affiliation{Instituto de Astrofísica de Canarias, Vía Láctea, s/n, E-38205 La Laguna, Spain}\affiliation{Departamento de Astrof\'\i sica, Universidad de La Laguna, E-38206 La Laguna, Spain}\affiliation{Spanish Space Solar Physics Consortium}\email{ckuckein@iac.es}

\author[orcid=0000-0002-3418-8449,sname='Solanki']{Sami~K.~Solanki} \affiliation{Max-Planck-Institut für Sonnensystemforschung, Justus-von-Liebig-Weg 3, 37077 Göttingen, Germany}\email{solanki@mps.mpg.de}

\author[orcid=0000-0003-1459-7074,sname='Lagg']{Andreas~Lagg} \affiliation{Max-Planck-Institut für Sonnensystemforschung, Justus-von-Liebig-Weg 3, 37077 Göttingen, Germany}\email{lagg@mps.mpg.de}
\author[orcid=0000-0002-9972-9840,sname='Gandorfer']{Achim~Gandorfer} \affiliation{Max-Planck-Institut für Sonnensystemforschung, Justus-von-Liebig-Weg 3, 37077 Göttingen, Germany}\email{gandorfer@mps.mpg.de}
\author[orcid=0000-0002-5054-8782,sname='Katsukawa']{Yukio~Katsukawa} \affiliation{National Astronomical Observatory of Japan, 2-21-1 Osawa, Mitaka, Tokyo 181-8588, Japan}\affiliation{Department of Astronomy, The University of Tokyo, 7-3-1, Hongo, Bunkyo-ku, Tokyo 113-0033, Japan}\affiliation{Department of Astronomical Science, The Graduate University for Advanced Studies (SOKENDAI), 2-21-1 Osawa, Mitaka, Tokyo 1818588, Japan}\email{yukio.katsukawa@nao.ac.jp}
\author[orcid=0000-0002-0787-8954,sname='Bernasconi']{Pietro~Bernasconi} \affiliation{Johns Hopkins University Applied Physics Laboratory, 11100 Johns Hopkins Road, Laurel, Maryland, USA}\email{pietro.bernasconi@jhuapl.edu}
\author[sname='Berkefeld']{Thomas~Berkefeld} \affiliation{Institut für Sonnenphysik (KIS), Georges-Köhler-Allee 401a, 79110 Freiburg, Germany}\email{thomas.berkefeld@leibniz-kis.de}
\author[orcid=0009-0009-4425-599X,sname='Feller']{Alex~Feller} \affiliation{Max-Planck-Institut für Sonnensystemforschung, Justus-von-Liebig-Weg 3, 37077 Göttingen, Germany}\email{feller@mps.mpg.de}
\author[orcid=0000-0001-6317-4380,sname='Riethmüller']{Tino~L.~Riethmüller} \affiliation{Max-Planck-Institut für Sonnensystemforschung, Justus-von-Liebig-Weg 3, 37077 Göttingen, Germany}\email{riethmueller@mps.mpg.de}


\author[orcid=0000-0001-5616-2808,sname='Kubo']{Masahito~Kubo} \affiliation{National Astronomical Observatory of Japan, 2-21-1 Osawa, Mitaka, Tokyo 181-8588, Japan}\email{masahito.kubo@nao.ac.jp}
\author[orcid=0000-0003-3490-6532,sname='Smitha']{H.~N.~Smitha} \affiliation{Max-Planck-Institut für Sonnensystemforschung, Justus-von-Liebig-Weg 3, 37077 Göttingen, Germany}\email{narayanamurthy@mps.mpg.de}
\author[sname='Grauf']{Bianca~Grauf} \affiliation{Max-Planck-Institut für Sonnensystemforschung, Justus-von-Liebig-Weg 3, 37077 Göttingen, Germany}\email{grauf@mps.mpg.de}
\author[sname='Carpenter']{Michael~Carpenter} \affiliation{Johns Hopkins University Applied Physics Laboratory, 11100 Johns Hopkins Road, Laurel, Maryland, USA}\email{michael.carpenter@jhuapl.edu}
\author[sname='Bell']{Alexander~Bell} \affiliation{Institut für Sonnenphysik (KIS), Georges-Köhler-Allee 401a, 79110 Freiburg, Germany}\email{albe@leibniz-kis.de}
\author[orcid=0000-0001-7764-6895,sname='Martínez~Pillet']{Valentín~Martínez~Pillet} \affiliation{Instituto de Astrofísica de Canarias, Vía Láctea, s/n, E-38205 La Laguna, Spain}\affiliation{Departamento de Astrof\'\i sica, Universidad de La Laguna, E-38206 La Laguna, Spain}\affiliation{Spanish Space Solar Physics Consortium}\email{vmpillet@iac.es}


\author[orcid=0000-0003-4319-2009,sname='Castellanos~Durán']{Juan~Sebastián~Castellanos~Durán} \affiliation{Max-Planck-Institut für Sonnensystemforschung, Justus-von-Liebig-Weg 3, 37077 Göttingen, Germany}\email{castellanos@mps.mpg.de}
\author[orcid=0009-0002-6808-5154,sname='Harnes']{Edvarda~Harnes} \affiliation{Max-Planck-Institut für Sonnensystemforschung, Justus-von-Liebig-Weg 3, 37077 Göttingen, Germany}\email{harnes@mps.mpg.de}
\author[orcid=0000-0001-6029-7529,sname='Hoelken']{Johannes~Hoelken} \affiliation{Max-Planck-Institut für Sonnensystemforschung, Justus-von-Liebig-Weg 3, 37077 Göttingen, Germany}\email{hoelken@mps.mpg.de}
\author[orcid=0000-0003-1409-1145,sname='Iglesias']{Francisco~A.~Iglesias} \affiliation{Max-Planck-Institut für Sonnensystemforschung, Justus-von-Liebig-Weg 3, 37077 Göttingen, Germany}\affiliation{Grupo de Estudios en Heliofísica de Mendoza, CONICET, Universidad de Mendoza, Boulogne sur Mer 683, 5500 Mendoza, Argentina}\email{iglesias@mps.mpg.de}
\author[orcid=0000-0002-4669-5376,sname='Ishikawa']{Ryohtaroh~T.~Ishikawa} \affiliation{National Institute for Fusion Science, 322-6 Oroshi-cho, Toki City 509-5292, Japan}\email{ishikawa.ryohtaro@nifs.ac.jp}
\author[orcid=0000-0001-7452-0656,sname='Kawabata']{Yusuke~Kawabata} \affiliation{National Astronomical Observatory of Japan, 2-21-1 Osawa, Mitaka, Tokyo 181-8588, Japan}\email{kawabata.yusuke@nao.ac.jp}
\author[orcid=0000-0002-1043-9944,sname='Matsumoto']{Takuma~Matsumoto} \affiliation{Centre for Integrated Data Science, Institute for Space-Earth Environmental Research, Nagoya University, Furocho, Chikusa-ku, Nagoya, Aichi 464-8601, Japan}\email{takuma.matsumoto@gmail.com}
\author[orcid=0000-0002-7044-6281,sname='Oba']{Takayoshi~Oba} \affiliation{Advanced Research Center for Space Science and Technology, Institute of Science and Engineering, Kanazawa University, Kakuma-machi, Kanazawa, Ishikawa 920-1192, Japan}\affiliation{Max-Planck-Institut für Sonnensystemforschung, Justus-von-Liebig-Weg 3, 37077 Göttingen, Germany}\email{oba@mps.mpg.de}
\author[orcid=0000-0003-1483-4535,sname='Strecker']{Hanna~Strecker} \affiliation{Instituto de Astrofísica de Andalucía, CSIC, Glorieta de la Astronomía s/n, 18008 Granada, Spain}\affiliation{Spanish Space Solar Physics Consortium}\email{streckerh@iaa.es}
\author[orcid=0000-0003-1971-5551,sname='Vukadinović']{Dušan~Vukadinović} \affiliation{Institut für Physik, Universität Graz, Universitätsplatz 5, 8010 Graz, Austria}\affiliation{Max-Planck-Institut für Sonnensystemforschung, Justus-von-Liebig-Weg 3, 37077 Göttingen, Germany}\email{vukadinovic@mps.mpg.de}


\begin{abstract}
A part of the magnetic energy released during a flare is transported to the lower atmosphere. High-resolution observations show that flare ribbons, sites of energy deposition at the footpoints of flaring loops which appear bright in the chromosphere and transition region, are structured on small spatial scales on the order of 100\,km. Based on idealized numerical models of flares it is suggested that the ribbon fine-structures could originate from a tearing instability and the development of plasmoids in current sheets. Here we report on Fe\,{\sc i}\,5250.6\,\AA\ and Mg\,{\sc i}\,b$_2$\,5173\,\AA\ spectral observations of a solar flare from the Tunable Magnetograph onboard the \sunrise{} balloon-borne mission that reveal an intricate link between the flare ribbon structure and the ambient chromosphere. We identified uncombed chromospheric loops and non-flaring fine-structures that are interspersed among brighter flare ribbon threads. These loops remain stable on timescales of minutes. Spectral lines from these regions show reduced emission or self-reversal in the line core compared with the immediately adjacent flare ribbons. We discuss the potential role of these structures in the onset of a flare. Furthermore, we suggest that irrespective of the complexities in the flaring current sheet, uncombed chromospheric loops and nonflaring fine-structure might play a role in spatially modulating the flare energy deposition in the lower atmosphere.       
\end{abstract}


\section{Introduction}
Flares are major disturbances in the solar atmosphere powered by a rapid conversion of magnetic energy into intense plasma heating, mass ejections, waves, and particles accelerated to high energies \citep{2002A&ARv..10..313P,2011LRSP....8....6S,2017LRSP...14....2B}. High-resolution coronal extreme ultraviolet observations showcase substantial spatial (down to a few 100\,km) and temporal (down to 2\,s) structuring in solar flares, offering a glimpse into the complex magnetic energy release process \citep[][]{2025SoPh..300..152R,2026A&A...705A.113C}.

During a flare, energy from a coronal reconnection site is transported into the lower atmosphere to the footpoints of flaring loops, giving rise to distinct kernels which appear bright over a wide range of wavelengths from visible to hard X-rays \citep[][]{2013ApJ...766..127Y,2024A&A...692A.176C,2025ApJ...986...73L,2025ApJ...990L...3T}. The first flare kernels form in concurrence with the formation of reconnection-prone hot flux rope in the corona \citep[][]{2021ApJ...911..133C}. Groups of kernels forming flare ribbons exhibit a high degree of spatial structuring \citep[][]{2013ApJ...771..104F,2024A&A...685A.137P}. Individual kernels have cross-sectional sizes on the order of 100\,km \citep[][]{2016NatSR...624319J}, and are observed to be separated by distances of about 500\,km \citep[][]{2025A&A...693A...8T}. The sizes and spatial separation of flare kernels are likely limited by the resolution of the telescope.

\begin{figure*}
 \begin{center}
   \includegraphics[width=\textwidth]{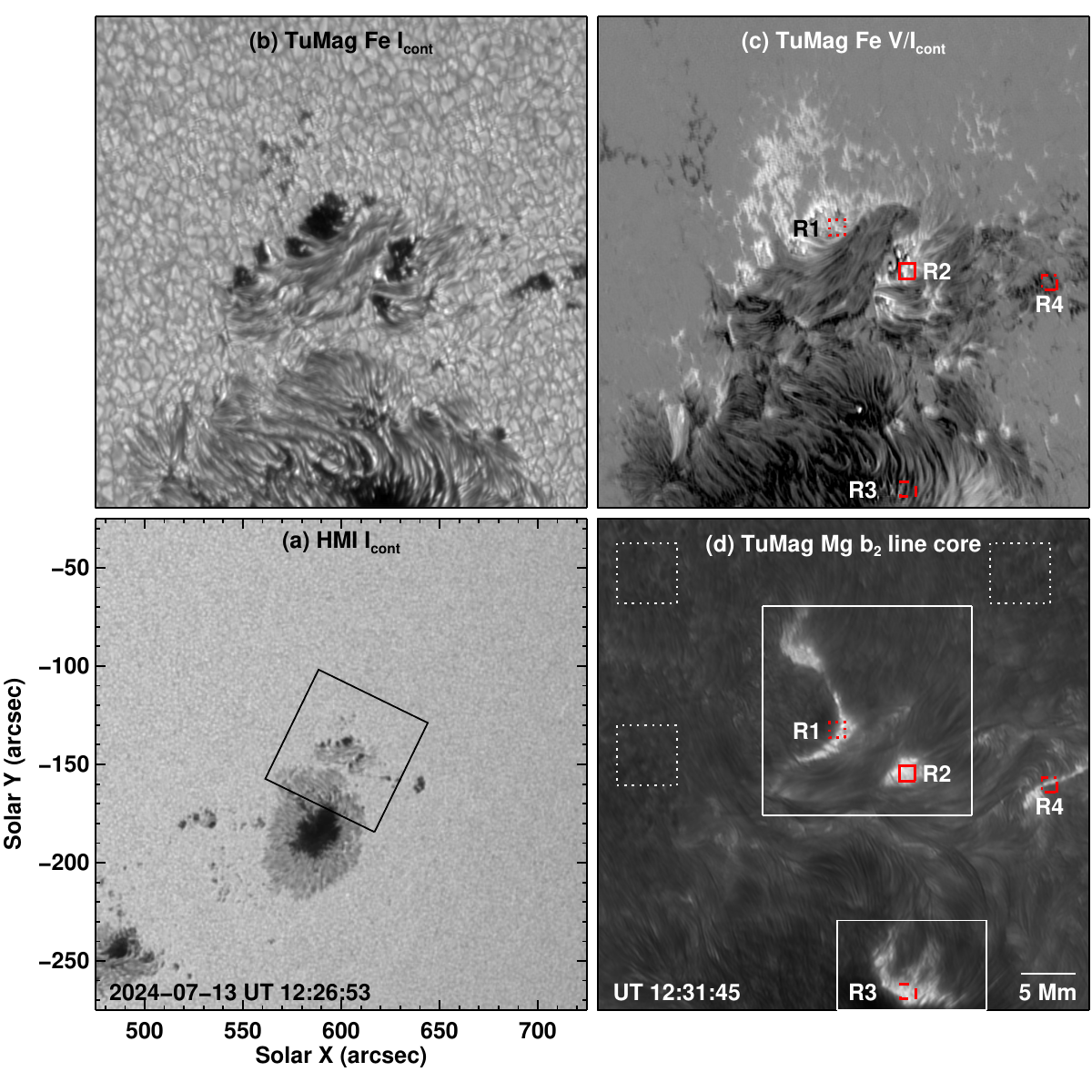}
   \caption{Lower atmospheric overview of the M5 class flare on 2024 July 13 UT\,12:30. (a) SDO/HMI continuum intensity map of AR13738. North is oriented up. The slanted black box is the TuMag field of view covering a section of the leading polarity of active region. (b) TuMag continuum intensity map at 0.227\,\AA\ away from the Fe\,{\sc i}\,5250.6\,\AA\ line. (c) TuMag blue-wing Stokes $V$ signal, normalized to the continuum intensity. The lighter (darker) shaded pixels represent regions with positive (negative) polarity magnetic field. (d) Map of the Mg\,{\sc i}\,b$_2$\,5173\,\AA\ line core recorded by TuMag. The three white dotted boxes represent quiet-Sun regions. The red boxes (R1--R4), also marked in panel c, are overlaid on the four flare ribbons. The northern and southern ribbon (outlined by a white square and rectangle, respectively) are further shown in Figs.\,\ref{fig:north} and \ref{fig:south}. There is an animation associated with panel (d) in the online Journal. The animation has a play back time of 6\,s\ and contains 59 frames with time stamps from 2024\,July\,13\,UT\,11:44 to UT\,12:55, with $\sim$73.5\,s increments. The frames in the movie are contrast enhanced using a multi-Gaussian normalization technique \citep[][]{2014SoPh..289.2945M}.\label{fig:over}}
 \end{center}
\end{figure*}

Probing the structure of flare ribbons and their evolution is providing insights into the nature of magnetic reconnection and energy release processes in a flare \citep[][]{2016ApJ...823...41D,2017ApJ...845...49K,2022SoPh..297...80Q,2023ApJ...958..104K}. For instance, reconnection rates as inferred from the photospheric magnetic field based on the spatial development of flare ribbons, and the hard X-ray emission show bursts of near co-temporal quasi-periodic pulsations (QPP). This coincidence hints at an oscillatory process and plasmoid dynamics in the reconnection region \citep[][]{2024ApJ...965...16C}. Such QPPs in hard X-ray emission are also shown to correlate with the fluctuations in the condensation downflows in the chromosphere, suggestive of a repeated reconnection driver \citep[][]{2026NatAs..10...54A}. Spatial separation of flare kernels and their further complex evolution is thought to be a result of tearing mode instability \citep[][]{2025A&A...693A...8T,2025ApJ...995L..54F}. 

Numerical simulations have advanced greatly in the recent years to capture the nature of energy conversion and transport in flaring loops \citep[][]{2023ApJ...947...67R,2024A&A...684A.171D}. Three-dimensional (3D) simulations of highly sheared and eruptive flux ropes do qualitatively reproduce the spatial structure of flare ribbons and the undulations that they undergo \citep[][]{2021ApJ...920..102W,2025ApJ...993...31D}. The ribbon structures are ascribed to the plasmoids and 3D flux ropes that are generated in a flare current sheet by reconnection. Given that these simulations do not include the magnetically highly structured lower atmosphere at the coronal base, where the flare ribbon signatures are typically observed, a direct correspondence between observations and simulations is not fully established. In this study we present an alternative explanation to the structure of the flare ribbon in the lower atmosphere with high-resolution observations from the \sunrise{} balloon-bourne observatory \citep[][]{2010ApJ...723L.127S,2017ApJS..229....2S,solankietal2026,2011SoPh..268....1B,2025SoPh..300...75K}.

\begin{figure}
 \begin{center}
   \includegraphics[width=0.4725\textwidth]{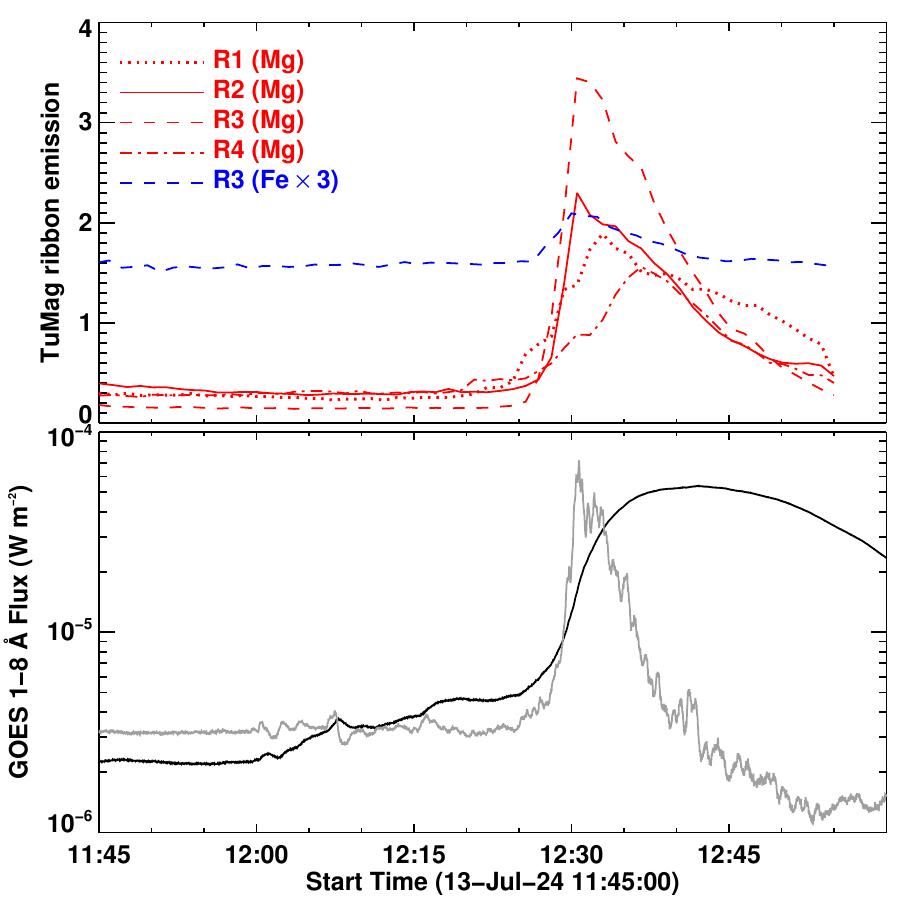}
   \caption{Evolution of the flare emission. Upper panel: Spatially averaged Mg\,{\sc i}\,b$_2$ line core intensities, normalized to the local continuum intensities, from the four ribbon locations (red boxes R1--R4 in Fig.\,\ref{fig:over}d) are plotted as functions of time with corresponding red colored line styles. The blue curve is the Fe\,{\sc i}\,5250.6\,\AA\ line red-wing intensity normalized to the local continuum at ribbon location R3. Lower panel: Time series of the disk-integrated GOES X-ray flux (black) and the flux derivative (gray). \label{fig:lc}}
 \end{center}
\end{figure}

\begin{figure}
 \begin{center}
   \includegraphics[width=0.4725\textwidth]{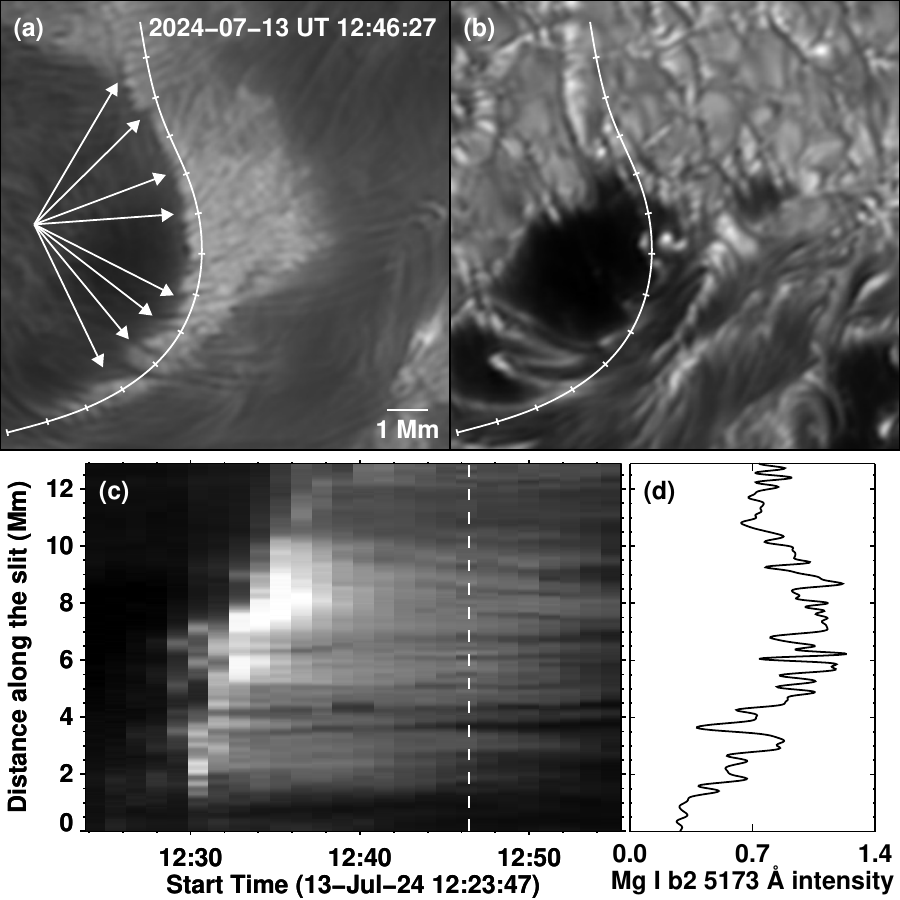}
   \caption{Spatial characteristics of uncombed chromospheric loops. (a) Zoomed-in field of view of the elongated northern ribbon site, that encloses the dotted red box in Fig.\,\ref{fig:over}. The white curve, with 1\,Mm tick marks, cuts through the ribbon emission. Arrows point to examples of uncombed chromospheric loops interspersed among the flare ribbon. (b) The corresponding continuum map obtained at $+650$\,m\AA\ away from the Mg\,{\sc i}\,b$_2$ line core. (c) Intensity interpolated along the white curve is stacked and displayed as a function of time. (d) Mg\,{\sc i}\,b$_2$ line core intensity along the white curve at the time-stamp marked by the dashed line in panel c. \label{fig:ribbon}}
 \end{center}
\end{figure}

\begin{figure*}
 \begin{center}
   \includegraphics[width=\textwidth]{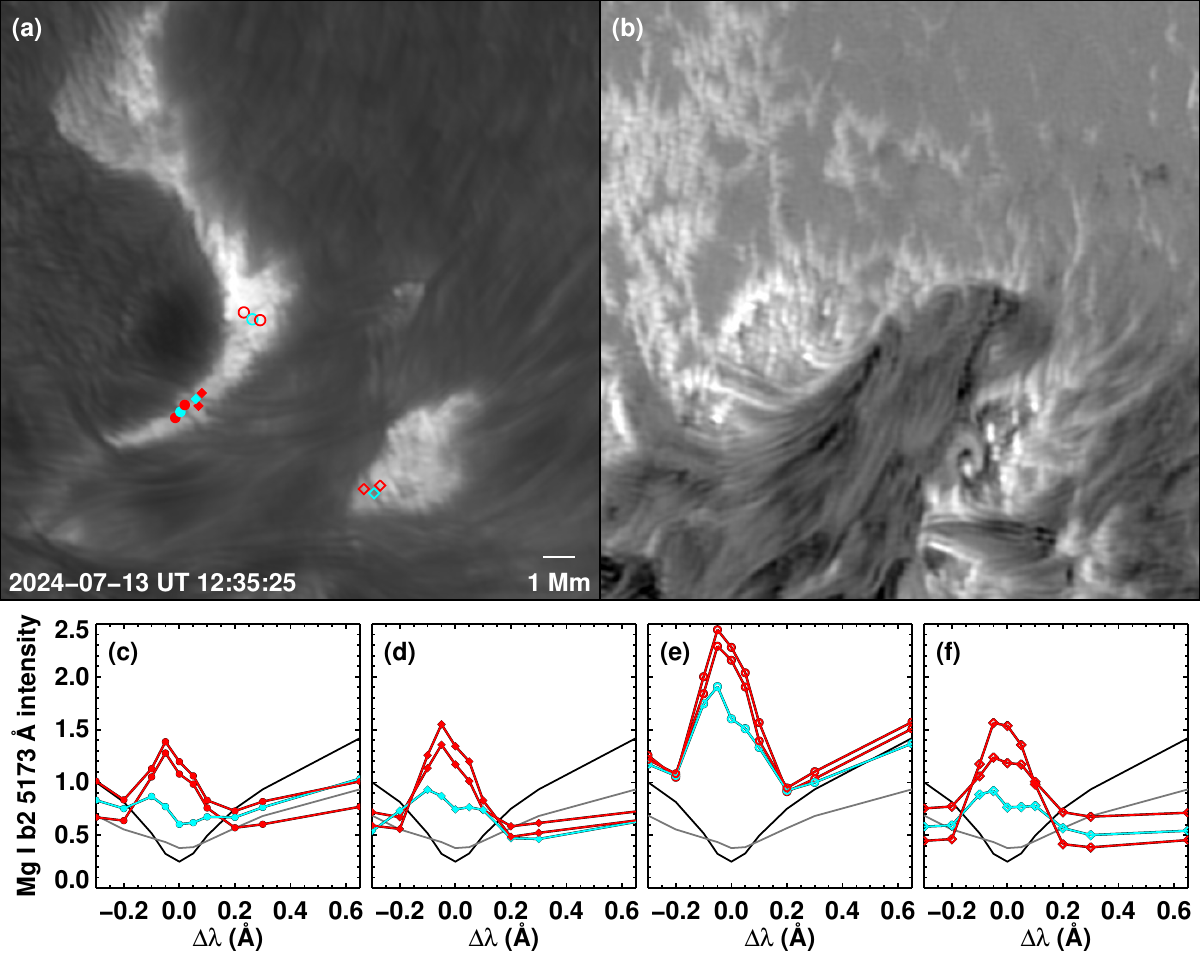}
   \caption{Chromospheric and magnetic substructure at the site of flare energy deposition. (a) Zoom into the northern ribbon region. This field of view is also outlined by a white square in Fig.\,\ref{fig:over}d. The circle and diamond symbols are overlaid on different Mg\,{\sc i}\,b$_2$ line core features. The red colored symbols overlie the locations that exhibit signatures of flare energy deposition. The cyan symbols overlie the uncombed chromospheric loops, flanked by the brighter ribbon features on either sides. (b) Photospheric Fe\,{\sc i} continuum-normalized Stokes $V$ map showing complex structure of the magnetic field. Mg\,{\sc i}\,b$_2$ line profiles as a function of wavelength (relative to the line core), with filled circles (panel c), filled diamonds (panel d), open circles (panel e), open diamonds (panel f), emerging from various corresponding regions identified in panel (a) are plotted. The black curve represents the quiet Sun profile of the Mg\,b$_2$ line averaged from the three white boxed regions in Fig.\,\ref{fig:over}d. The gray curve is the average line profile from a penumbral region devoid of flare ribbons. \label{fig:north}}
 \end{center}
\end{figure*}

\section{Observations\label{sec:obs}}
We present observations of an M5.3 class flare on 2024 July 13 recorded by the instruments on board the \sunrise{} observatory. The Tunable Magnetograph for the \sunrise{} mission \citep[TuMag;][]{2025SoPh..300..148D} obtained spectropolarimetric observations of the Fe\,{\sc i}\,5250.6\,\AA\ and the Mg\,{\sc i}\,b$_2$\,5173\,\AA\ line. The Fe\,{\sc i} line is sampled over seven wavelength points in addition to a continuum point located at $+227$\,m\AA\ away from its line core. The Mg\,{\sc i}\,b$_2$ line is sampled at nine wavelength points along with a continuum wavelength point located at $+650$\,m\AA\ away from its line core. Under quiet-Sun conditions, the core of this Mg line samples low chromospheric heights of about 700-900\,km above the solar surface \citep[][]{1988ApJ...330.1008M,2011A&A...531A..17R,2025A&A...696A.105S,2025A&A...696A.106S}. These TuMag data have an image scale of 0.037$''$\,pixel$^{-1}$ (diffraction-limited resolution of 0.13$''$ equivalent to about 95\,km\ at the disk center), and an average cadence of $\sim$73.5\,s. The data were reduced using the TuMag data reduction pipeline\footnote{TuMag's pipeline is publicly available at \url{https://github.com/PabloSGN/TuMags_Reduction_Pipeline}} (Orozco Su\'arez et al. 2026, in prep.).
For our analysis, we make use of the phase-diversity reconstructed \citep[e.g.,][]{2022ApJS..263....7B} Stokes $I$ and $V$ maps covering a period of 70\,minutes, starting at UT\,11:44. The Stokes $V$ data presented here should be considered preliminary as there could still be some residual crosstalk from Stokes $I$ into $V$, which might not be uniform across the field of view.

For additional context, we use photospheric continuum intensity maps from the Helioseismic and Magnetic Imager \citep[HMI;][]{2012SoPh..275..207S} on board the Solar Dynamics Observatory \citep[SDO;][]{2012SoPh..275....3P}. We complemented these with coronal EUV images from the 193\,\AA\ passband of the Atmospheric Imaging Assembly \citep[AIA;][]{2012SoPh..275...17L} onboard SDO. These HMI and AIA data are processed through the standard packages available in solarsoft \citep[][]{1998SoPh..182..497F}. To evaluate chromospheric structures prior to and after the flare, we make use of data from the Chinese H$\alpha$ Solar Explorer \citep[CHASE;][]{2022SCPMA..6589602L,2022SCPMA..6589603Q} in the time period UT\,11:00 to UT\,13:00 on 2024 July 13. These data have an image scale of 1$''$ and cadence of 72\,s. The CHASE data between UT\,11:28 and UT\,12:37 are not available.

\section{Overview of the flaring region\label{sec:over}}

The flare occurred in an active region, with the National Oceanic and Atmospheric Administration (NOAA) number 13738, around the northern periphery of its leading sunspot (Helioprojective Cartesian coordinates of  $x=+600''$ and $y=-150''$). The SDO/HMI map of the region shows a group of pores and penumbral filaments adjacent to a larger sunspot (Fig.\,\ref{fig:over}a). The fine structure of the region is better seen in the high-resolution TuMag data. The highly sheared nature of penumbral structures is evident in the photospheric continuum intensity (Fig.\,\ref{fig:over}b). The Stokes $V$ map from the blue wing of the Fe\,{\sc }, normalized to the continuum intensity, points to a highly filamentary nature of the surface magnetic field in that region. There are minority polarity magnetic patches scattered within the penumbral structures (Fig.\,\ref{fig:over}c). The system of pores adjacent to the sheared penumbral region in the center of the field of view have opposite magnetic polarity to that of the larger sunspot. In Fig.\,\ref{fig:over}d, we plot the TuMag Mg\,b$_2$ line core map retrieved, centered on the flaring region, with four main ribbon features (R1--R4) identified. While the pair R1-R2 overlies positive (minor) polarity magnetic field region, the ribbon structures R3--R4 overlie the dominant negative polarity patches. 

The disk-integrated soft X-ray observations recorded by GOES show that, after preflare brightening for about 30\,minutes, the flare started around UT\,12:26 and reached its peak flux around UT\,12:40. We found that ribbon features, in particular, R1, R3, and R4, exhibit noticeable enhancements in the Mg line core intensity starting around UT\,12:19:31 (Figure\,\ref{fig:over} animation). Thus the chromospheric response precedes the impulsive rise phase of the flare by at least 10\,minutes. Following this early preflare response, the Mg line core intensity from all the ribbon sites shows a rapid rise during the impulsive phase of the flare, which coincides well with the GOES flux derivative. The ribbon emission gradually decays over a period of about 20\,minutes afterwards (Fig.\,\ref{fig:lc}). Flare ribbon emission is also detected in the Fe\,{\sc i} line core and its wings (to at least $\pm$80\,m\AA\ away from the line core), particularly at locations R1, R2, and R3. We demonstrate this with a plot of the Fe\,{\sc i} line red-wing emission ($+$40\,m\AA\ away from the line core) at ribbon R3 in Fig.\,\ref{fig:lc}. Based on 1D radiative-hydrodynamic flare simulations, it is argued that the Fe\,{\sc i} line emission in flares could originate from dense chromospheric condensations \citep[][]{2024ApJ...963...40M}. Interestingly, there are identifiable, narrow regions devoid of brighter ribbon intensity at these sites. We will analyse the spatial characteristics and spectral properties of these features in the following. 

\section{Uncombed chromospheric loops within flare ribbons\label{sec:res}}

Consistent with recent high-resolution coronal and chromospheric studies, a closer examination of the flare ribbons in our observations reveals that they are not contiguous. There are well-defined narrow ribbon-free sites. We demonstrate this in Fig.\,\ref{fig:ribbon} in which we zoom into one of the (elongated) northern ribbon sites. Importantly, these regions host low-lying chromospheric loops that are apparently not as bright as the individual threads in the flare ribbons. We define these elongated low-lying loops, that cross the flare ribbon, as the uncombed chromospheric loops. Our nomenclature is analogous to the model of uncombed penumbral fields, in which horizontal magnetic flux tubes are embedded in the background inclined field \citep[][]{1993A&A...275..283S,2000A&A...361..734M}. Such an uncombed loop system with interlaced horizontal and vertical components of the magnetic field has also been inferred with the observations of chromospheric fibrils in the remnants of an active region \citep[][]{2011ApJ...742..119R}. We point to some of the clear examples of this type of structure in Fig.\,\ref{fig:ribbon}a. The orientation of these uncombed chromospheric loops is not aligned with the underlying highly sheared penumbral filaments which are bent northward (Fig.\,\ref{fig:ribbon}b). This points to variations in the magnetic field direction with height in this region, which would contribute to the magnetic complexity.

The spatial corrugation created by these uncombed chromospheric structures and their temporal stability are better seen through the Mg line core space-time map plotted along the white line that cuts through several of these features (Fig.\,\ref{fig:ribbon}c). Due to the enhanced emission in the Mg\,b$_2$ line core, the brighter regions in the space-time map correspond to the flare ribbons, separated on scales of a few 100\,km. In between we see troughs that are associated with the nonflaring uncombed loops (panel d). Despite the motion and general evolution of the nonflaring loops, we found that they remain nearly stable for a prolonged period at most of the locations among the ribbon site (Fig.\,\ref{fig:ribbon}c), which can be seen as elongated dark lanes in the space-time map. We will discuss further this aspect of their longevity and implications in Sect.\,\ref{sec:disc}. 

\begin{figure*}
 \begin{center}
   \includegraphics[width=0.49\textwidth]{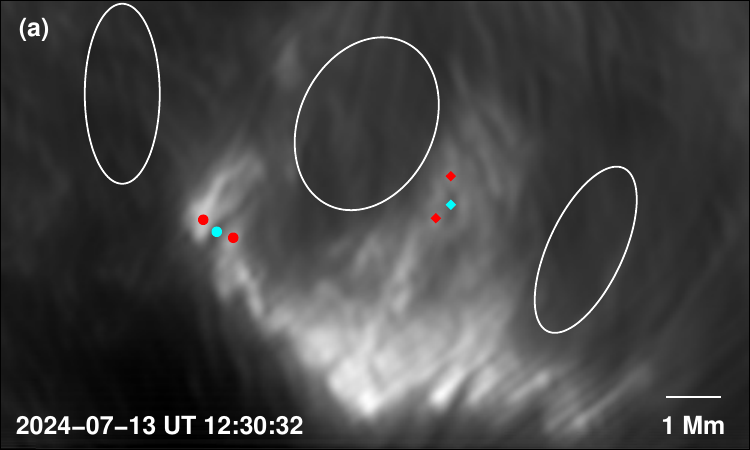}
   \includegraphics[width=0.49\textwidth]{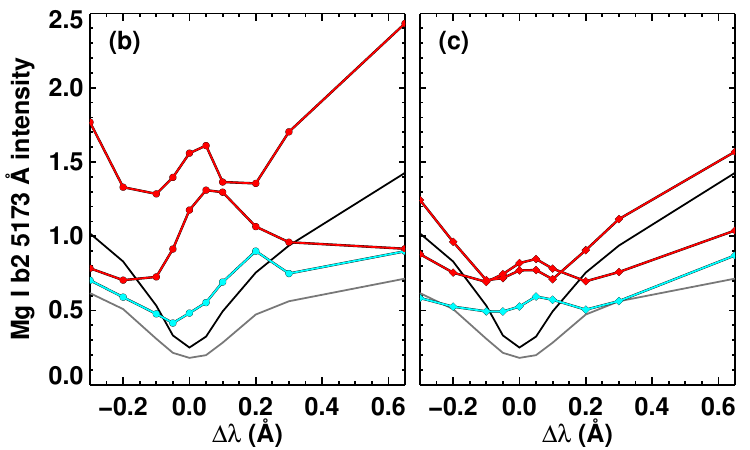}
   \caption{Similar to Figure\,\ref{fig:north}, but plotted for the southern ribbon site identified with the white rectangle in Fig.\,\ref{fig:over}d. The white ellipses enclose larger regions with no signatures of flare energy deposition.\label{fig:south}}
 \end{center}
\end{figure*}

To further characterize the nature of these nonflaring uncombed chromospheric loops, we examine their spectral profiles by exploiting the high-quality high-resolution TuMag data. In Fig.\,\ref{fig:north}a we show examples with groups of uncombed chromospheric loops and their immediate neighboring ribbon features. The spectral profiles are displayed in Fig.\,\ref{fig:north}b--e. The ribbon sites exhibit enhanced core emission, but in between a pair of these brighter features the embedded uncombed loops show overall reduced emission in the line core. There are also clear signatures of central reversals at these nonflaring sites. Similar to the northern ribbons, the southern ribbon site shows a distinction between flare ribbons and nonflaring fine-structures (Fig.\,\ref{fig:south}). We note that several of these spectral profiles show shallower wings than the quiet-Sun profile. This is mainly because the sites we selected directly overlie the penumbral filaments, which are in general darker than the quiet-Sun granulation. As such, the average spectral profile from the penumbral regions devoid of flare emission shows shallower wings than the quiet-Sun region. 

The spatial extent of each group of symbols covering a pair of ribbon threads and embedded uncombed chromospheric structures is less than 1\,Mm, which further emphasizes the small-scale nature of the nonflaring uncombed loops within the ribbon site. In addition to the existence of this fine-structure, we also identified larger regions on megameter scales that are devoid of flare energy deposition despite them being flanked by flare ribbons on either side (regions enclosed by ellipse in Fig.\,\ref{fig:south}a). Networks of stable chromospheric loops are rooted in these larger ribbon-free regions. Similar uncombed fine-structures are also observed to cross the flare ribbon R4 marked in Fig.\,\ref{fig:over}d. 

\begin{figure*}
 \begin{center}
   \includegraphics[width=\textwidth]{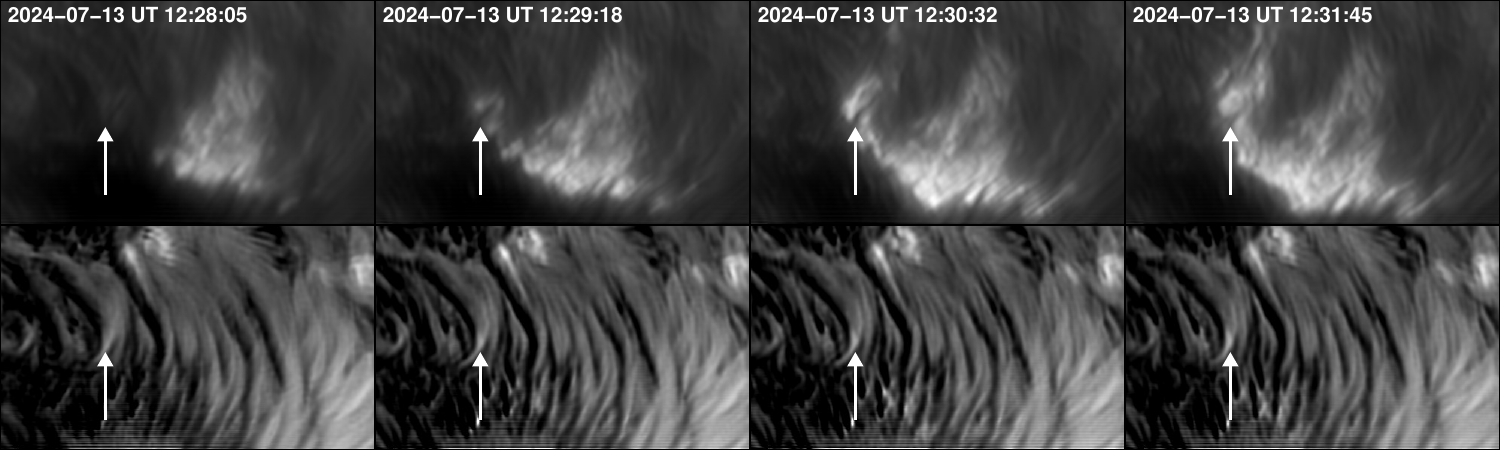}
   \caption{Subtle nature of the nonflaring fine-structure.  The field of view is the same as in Fig.\,\ref{fig:south}a. Sequences of Mg line core maps (upper panels) and Fe\,{\sc i} normalized Stokes $V$ maps (lower panels) are plotted. The arrow points to the chromospheric nonflaring fine-structure overlying a potential mixed-polarity or a twisted magnetic structure in a penumbral filament.\label{fig:split}}
 \end{center}
\end{figure*}

Our observations further reveal the impact of the lower atmospheric nonflaring fine-structure on the flare ribbon evolution. At the spatial locations of the spectral profiles plotted in Fig.\,\ref{fig:south}b, the nonflaring structure directly overlies a penumbral filament that is highly twisted. There are tentative signatures of mixed-polarity magnetic structures around the twisted part of the penumbral filament  (Fig.\,\ref{fig:split}). The flare ribbon pivots northward exactly at this penumbral filament, which remains stable for several minutes. This example clearly demonstrates that the flare ribbon evolution is marked by even the subtle variations in the lower atmospheric magnetic field.

Zooming out, in addition to the localized uncombed structures within the various ribbons that TuMag captured, CHASE H$\alpha$ images reveal a longer chromospheric filament that is connected to the system of positive polarity pores and the negative polarity sunspot (Fig.\,\ref{fig:over2}a). An imprint of this larger filament is also observed as a darker absorption feature in the AIA 193\,\AA\ map (Fig.\,\ref{fig:over2}b). TuMag Mg\,b$_2$ data capture the lower chromospheric counterpart of this filament in the form of highly dynamic elongated loops (better visualized in the Figure\,\ref{fig:over} animation).

Intriguingly, the main  flaring loop bundle (hotter flux rope) in this active region shares its footpoints with the cooler filament, which is also rooted in the sunspot/pore group. Our observations thus point to a system of entwined cooler and hotter structures in the flaring region. A similar configuration is observed at the ribbon location R4 where low-lying filamentary structures in the TuMag data share a common footpoint with the hotter flux rope seen in the AIA 193\,\AA\ images.

\section{Discussion and Conclusion\label{sec:disc}}
\begin{figure*}
 \begin{center}
   \includegraphics[width=\textwidth]{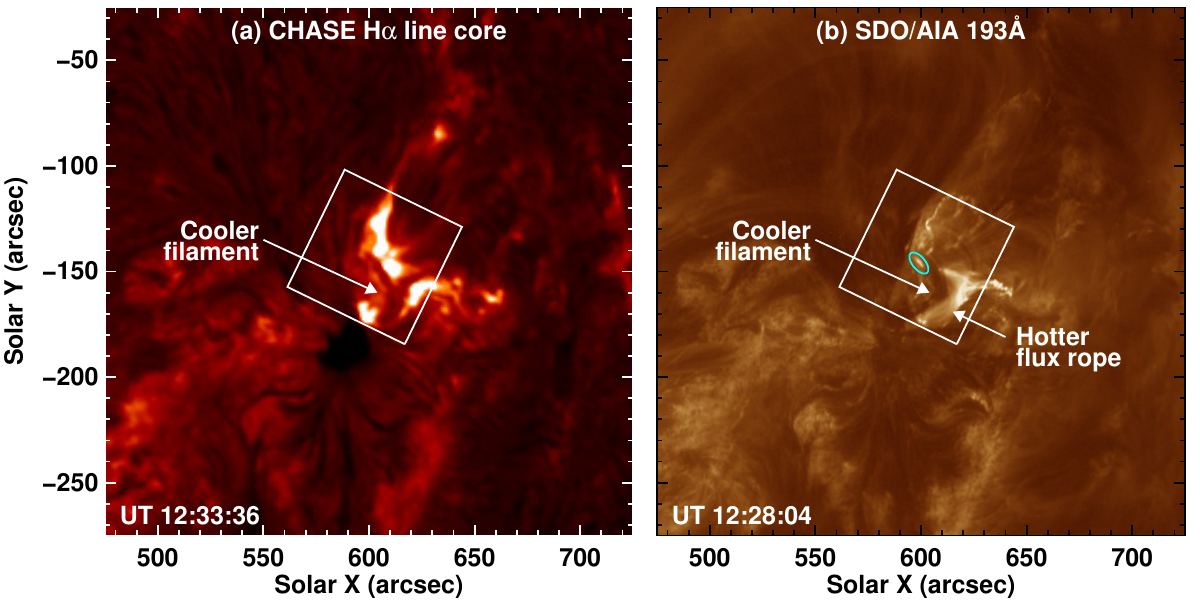}
   \caption{Chromospheric and coronal overview of the flaring region. (a) CHASE H$\alpha$ line core intensity covering the same field of view as in Fig.\,\ref{fig:over}a. (b) Map of the coronal emission recorded by the SDO/AIA 193\,\AA\ passband on a square-root intensity scale. In both panels the white square outlines the field of view of the TuMag map shown in Fig.\,\ref{fig:over}b--d. The cooler filament and hotter flux rope features are marked. The cyan ellipse encloses the footpoints of the hot flux rope that are wrapped underneath the cooler filament. North is oriented up. There is an animation associated with panel (a) in the online Journal. The animation has a play back time of 4\,s\ with time stamps from 2024\,July\,13\,UT\,11:01 to UT\,13:00, with $\sim$72\,s increments.\label{fig:over2}}
 \end{center}
\end{figure*}

The main result of our study is that some of the spatial structure and corrugations in the flare ribbons arise from the presence of uncombed chromospheric loops. By comparing the photospheric Stokes $V$ and the Mg\,b$_2$ line core images, we infer that the uncombed loops associated with ribbons R1 and R2 have one footpoint in the positive polarity pores and the other in the neighboring sheared penumbral arcade of negative polarity. Similarly, the loops from the R3 region 
extend to various scattered positive polarity pores in the field of view. Moreover, the uncombed chromospheric loops and fine-structure that we observed remain distinguishable even through the preflare phase (i.e., prior to UT\,12:26; Movie S1). This suggests that the cooler structures sampled by the Mg\,{\sc i}\,b$_2$ line are not associated with flare-driven rain, a phenomenon that is usually seen after the impulsive phase of the flare with post-flare loops cooling \citep[][]{2016NatSR...624319J,2025A&A...699A.121K}. As discussed in the previous section, these uncombed loops are at the footpoints of a longer filament in that active region. One end of the hotter flux rope is wrapped underneath the filament (blue ellipse region in Fig.\,\ref{fig:over2}b). Owing to its higher temperature, this hotter flux rope itself will not be sampled by the Mg\,{\sc i}\,b$_2$ line. Instead, we observe its footpoints in the form of flare ribbons.

Despite sharing its footpoints with the hotter flux rope that flared, the cooler filament is nearly undisturbed during this M-class event \citep[partly or completely undisturbed filaments in flaring regions are rather common, e.g.,][]{2022ApJ...940L..12H,2025A&A...699A.121K}. Its longevity is also reflected in the relatively stable nonflaring uncombed chromospheric loops and fine-structures that TuMag captured (Fig.\,\ref{fig:ribbon}c), although the immediately adjacent ribbon structures exhibited intermittent evolution. This implies that there are locally stable and unstable zones within the same flaring system forming over a polarity inversion line. It is possible that some of this stable magnetic structure, which is observed as uncombed chromospheric loops, might eventually flare up.

\begin{figure}
 \begin{center}
   \includegraphics[width=0.4725\textwidth]{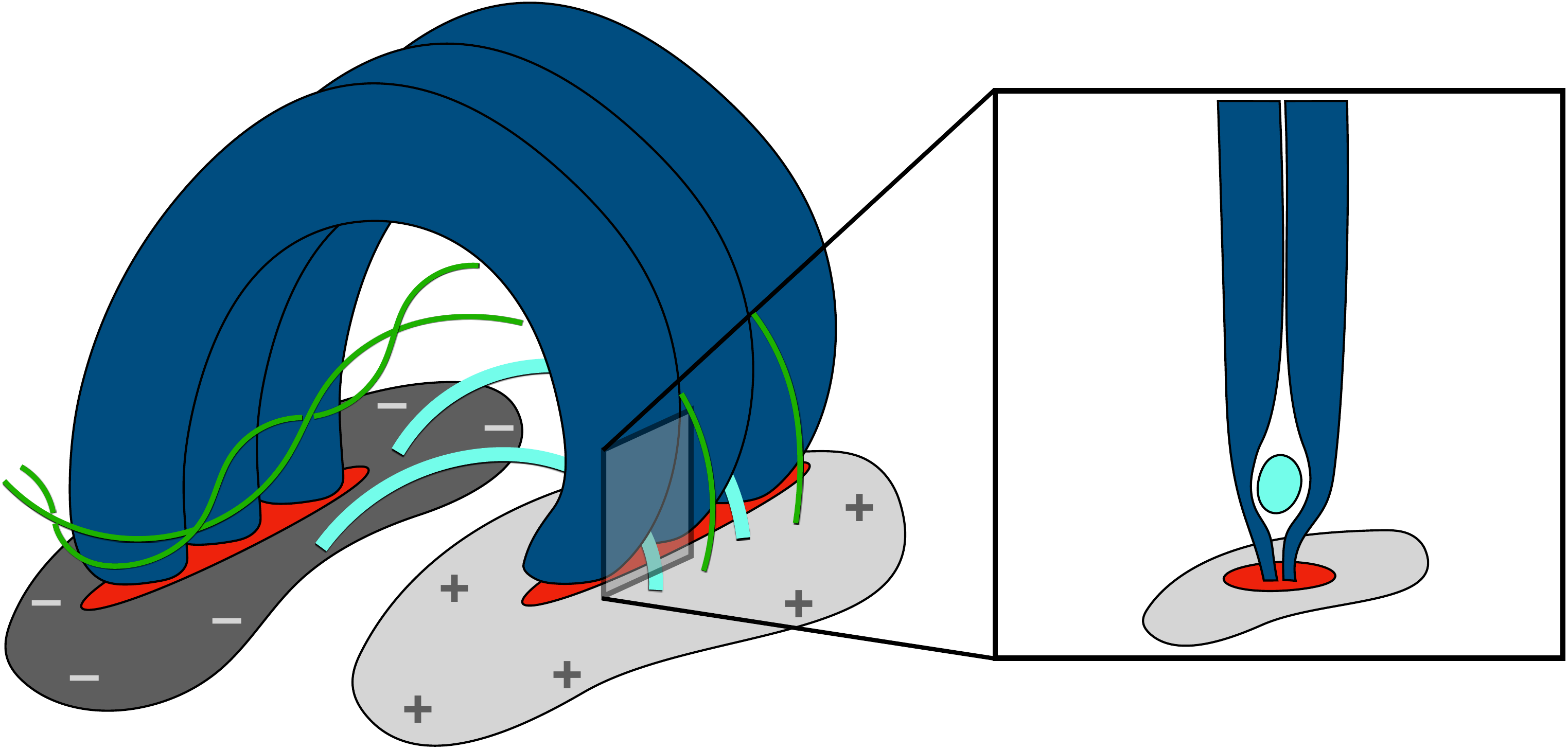}
   \caption{Simplified illustration of the spatial configuration of the uncombed chromospheric loops within the flare. Flaring loops (blue) and their footpoints indicative of ribbons (red), connected to a pair of opposite polarity magnetic field patches (regions labeled $+$ and $-$), are depicted. The lower-lying cyan colored loops indicate nonflaring uncombed chromospheric loops that are interspersed with the flare ribbons in the lower atmosphere. A pair of green-colored entwined field lines represent nonflaring cooler filament structures, which play the same role as the uncombed loops, but are connected elsewhere in the active region. The inset shows the side (in-plane) view of this configuration closer to the footpoints. The cyan shaded region represents the cross-section of an uncombed chromospheric loop with its axis oriented into the plane.    \label{fig:illus}}
 \end{center}
\end{figure}

We illustrate this scenario in Fig.\,\ref{fig:illus}. The uncombed chromospheric loops crossing flare ribbons could close down across the local polarity inversion line. Additionally, some of this structure could be threaded into a longer filament that connects to farther locations within the active region as is the case in this example. In our scenario, the shorter uncombed chromospheric loops and the longer filaments play the same role. Depending on the line of sight, these uncombed chromospheric loops, that are essentially cooler, would show reduced core emission in comparison to the heated flare ribbon. The flaring loops as they approach the surface would be wrapped around this nonflaring fine-structure, which would act as an absorber of the radiation coming from below. This is visible in the line profile as part of the emission from deeper layers producing an emission core in the Mg\,{\sc i}\,b$_2$. This emission is then partly absorbed by the cool loops in higher layers, leaving peaks on either side of the line core which exhibits a central reversal. Because of their low-lying nature, the uncombed loops might remain either darker or not as bright as the adjacent ribbon threads owing to limited flare energy deposition at those locations. As such, the presence of cooler structures in the foreground (e.g., surges), can obscure intense chromospheric emission from flare ribbons \citep[][]{2022A&A...659A..58P,2024A&A...688L...9C}.

Central reversals of chromospheric lines during flares are typically observed at the ribbon fronts during the impulsive phase. These spectral features spatially and temporally coincide with the nonthermal hard X-ray signatures \citep[][]{2017ApJ...842...82R,2018ApJ...861...62P,2023ApJ...944..104P}. Using higher resolution observations here we find that the central reversals, at least in the observed flare, are linked to uncombed chromospheric loops and nonflaring fine structures within flare ribbons. Interestingly, through a cursory inspection of the H$\alpha$ blue-wing and core data from the observations of \citet[][]{2016NatSR...624319J}, we noticed similar nonflaring structures persisting for several minutes as the flare ribbons sweep through. This hints at a prevalence of uncombed chromospheric loops in the flaring region. Therefore, the question, namely, how do these nonflaring structures resist intense heating, for instance through high-energy particle bombardment, remains open.

But the very existence of the uncombed chromospheric loops and the nonflaring fine-structures imaged by TuMag and their connections to the entwined cooler filament and hotter flux rope already during the preflare phase (Fig.\,\ref{fig:over2}b) has important implications. The temperature gradients arising from the uncombed loops or their filament extensions could lead to pressure perturbations across the hotter flux rope, triggering spatially and temporally discrete reconnection leading to intense localized heating events \citep[][]{2026A&A...705A.113C}. This is similar to the observations showing rapid coronal heating at the sites of chromospheric filaments \citep[][]{2022A&A...667A.166C}. This will lead to further destabilization of the flux rope, driving the flare \citep[][]{1999ApJ...510..485A,2001ApJ...552..833M,2026A&A...705A.113C}. 

Current high-resolution models of flares predict a continuous ribbon structure which is modulated by transient spiral structures which are traced back to plasmoid formation in the flaring current sheet \citep[][]{2025ApJ...993...31D}. We found that subtle variations in the lower atmospheric magnetic field can directly impact the progression of flare ribbons at those locations, by diverting their path (Fig.\,\ref{fig:split}). Such rapid directional changes at the fronts of flare ribbons could give the misimpression that they are undergoing spiral or swirling motions. Our observations thus suggest that the structure of flare ribbons could be influenced by the relatively stable uncombed loops and nonflaring fine-structures in the chromosphere. 

Flare ribbons being the sites of energy deposition, we suggest that their structuring could be governed by the complex fine-scale magnetic field of the chromosphere. Future simulations will have to incorporate the chromospheric complexity in order to assess coupling between various reconnection dynamics in flaring current sheets and their imprints in the lower atmosphere.

\begin{acknowledgements}
We thank the two anonymous referees for constructive comments  that helped us improved the presentation of the manuscript. This project has received funding from the European Research Council (ERC) under the European Union's Horizon Europe research and innovation programme (grant agreement Nos. 10103984 -- project ORIGIN; 101097844 -- project WINSUN). ALST acknowledges funding from the Consejer\'ia de Transformaci\'on Econ\'omica, Industria, Conocimiento y Universidades of the Junta de Andaluc\'ia through grant POSTDOC-21-00832. Sunrise III is supported by funding from the Max-Planck-Förderstiftung (Max Planck Foundation), NASA under Grants \#80NSSC18K0934 and \#80NSSC24M0024 (``Heliophysics Low Cost Access to Space'' program), and the ISAS/JAXA Small Mission-of-Opportunity program and JSPS KAKENHI Grant Numbers JP18H05234 and JP23K25916. This research has received financial support from the European Union’s Horizon 2020 research and innovation program under grant agreement No. 824135 (SOLARNET). It has also been funded by the Deutsches Zentrum für Luft- und Raumfahrt e.V. (DLR, grant no. 50 OO 1608). The Spanish contributions have been funded by the Spanish MCIN/AEI under projects RTI2018-096886-B-C5, and PID2021-125325OB-C5, and from ``Center of Excellence Severo Ochoa'' awards to IAA-CSIC (SEV-2017-0709, CEX2021-001131-S), all co-funded by European REDEF funds, ``A way of making Europe''. SDO is the first mission to be launched for NASA's Living With a Star (LWS) Program and the data supplied courtesy of the HMI and AIA consortia. CHASE mission is supported by China National Space Administration. We thank GOES team for making the X-ray data publicly available. C.K. acknowledges grant RYC2022-037660-I funded by MCIN/AEI/10.13039/501100011033 and by ``ESF Investing in your future'' and grant PID2024-156066OB-C55, funded by MCIN/AEI/ 10.13039/501100011033 and by ``ERDF A way of making Europe''.  
\end{acknowledgements}

\bibliographystyle{aasjournalv7}

\begin{thebibliography}{}
\expandafter\ifx\csname natexlab\endcsname\relax\def\natexlab#1{#1}\fi
\providecommand{\url}[1]{\href{#1}{#1}}
\providecommand{\dodoi}[1]{doi:~\href{http://doi.org/#1}{\nolinkurl{#1}}}
\providecommand{\doeprint}[1]{\href{http://ascl.net/#1}{\nolinkurl{http://ascl.net/#1}}}
\providecommand{\doarXiv}[1]{\href{https://arxiv.org/abs/#1}{\nolinkurl{https://arxiv.org/abs/#1}}}

\bibitem[{S.~K. {Antiochos} {et~al.}(1999){Antiochos}, {DeVore}, \&
  {Klimchuk}}]{1999ApJ...510..485A}
{Antiochos}, S.~K., {DeVore}, C.~R., \& {Klimchuk}, J.~A. 1999,
  \bibinfo{title}{{A Model for Solar Coronal Mass Ejections},} \apj, 510, 485,
  \dodoi{10.1086/306563}

\bibitem[{W. {Ashfield} {et~al.}(2026){Ashfield}, {Polito},
  {L{\"o}rin{\v{c}}{\'\i}k}, {De Pontieu}, {Chintzoglou}, {Bose}, {Freij},
  {Rouppe van der Voort}, {Joshi}, \& {Thoen Faber}}]{2026NatAs..10...54A}
{Ashfield}, W., {Polito}, V., {L{\"o}rin{\v{c}}{\'\i}k}, J., {et~al.} 2026,
  \bibinfo{title}{{Spectroscopic observations of solar flare pulsations driven
  by oscillatory magnetic reconnection},} Nature Astronomy, 10, 54,
  \dodoi{10.1038/s41550-025-02706-4}

\bibitem[{F.~J. {Bail{\'e}n} {et~al.}(2022){Bail{\'e}n}, {Orozco Su{\'a}rez},
  {Blanco Rodr{\'\i}guez}, \& {del Toro Iniesta}}]{2022ApJS..263....7B}
{Bail{\'e}n}, F.~J., {Orozco Su{\'a}rez}, D., {Blanco Rodr{\'\i}guez}, J., \&
  {del Toro Iniesta}, J.~C. 2022, \bibinfo{title}{{Performance of Sequential
  Phase Diversity with Dynamical Solar Scenes},} \apjs, 263, 7,
  \dodoi{10.3847/1538-4365/ac966d}

\bibitem[{P. {Barthol} {et~al.}(2011){Barthol}, {Gandorfer}, {Solanki},
  {Sch{\"u}ssler}, {Chares}, {Curdt}, {Deutsch}, {Feller}, {Germerott},
  {Grauf}, {Heerlein}, {Hirzberger}, {Kolleck}, {Meller}, {M{\"u}ller},
  {Riethm{\"u}ller}, {Tomasch}, {Kn{\"o}lker}, {Lites}, {Card}, {Elmore},
  {Fox}, {Lecinski}, {Nelson}, {Summers}, {Watt}, {Mart{\'\i}nez Pillet},
  {Bonet}, {Schmidt}, {Berkefeld}, {Title}, {Domingo}, {Gasent Blesa}, {del
  Toro Iniesta}, {L{\'o}pez Jim{\'e}nez}, {{\'A}lvarez-Herrero},
  {Sabau-Graziati}, {Widani}, {Haberler}, {H{\"a}rtel}, {Kampf}, {Levin},
  {P{\'e}rez Grande}, {Sanz-Andr{\'e}s}, \& {Schmidt}}]{2011SoPh..268....1B}
{Barthol}, P., {Gandorfer}, A., {Solanki}, S.~K., {et~al.} 2011,
  \bibinfo{title}{{The Sunrise Mission},} \solphys, 268, 1,
  \dodoi{10.1007/s11207-010-9662-9}

\bibitem[{A.~O. {Benz}(2017){Benz}}]{2017LRSP...14....2B}
{Benz}, A.~O. 2017, \bibinfo{title}{{Flare Observations},} Living Reviews in
  Solar Physics, 14, 2, \dodoi{10.1007/s41116-016-0004-3}

\bibitem[{L.~P. {Chitta} {et~al.}(2024){Chitta}, {Hannah}, {Fletcher},
  {Hudson}, {Young}, {Krucker}, \& {Peter}}]{2024A&A...688L...9C}
{Chitta}, L.~P., {Hannah}, I.~G., {Fletcher}, L., {et~al.} 2024,
  \bibinfo{title}{{Hard X-rays from the deep solar atmosphere. An unusual UV
  burst with flare properties},} \aap, 688, L9,
  \dodoi{10.1051/0004-6361/202450866}

\bibitem[{L.~P. {Chitta} {et~al.}(2021){Chitta}, {Priest}, \&
  {Cheng}}]{2021ApJ...911..133C}
{Chitta}, L.~P., {Priest}, E.~R., \& {Cheng}, X. 2021, \bibinfo{title}{{From
  Formation to Disruption: Observing the Multiphase Evolution of a Solar Flare
  Current Sheet},} \apj, 911, 133, \dodoi{10.3847/1538-4357/abec4d}

\bibitem[{L.~P. {Chitta} {et~al.}(2022){Chitta}, {Peter}, {Parenti},
  {Berghmans}, {Auch{\`e}re}, {Solanki}, {Aznar Cuadrado}, {Sch{\"u}hle},
  {Teriaca}, {Mandal}, {Barczynski}, {Buchlin}, {Harra}, {Kraaikamp}, {Long},
  {Rodriguez}, {Schwanitz}, {Smith}, {Verbeeck}, {Zhukov}, {Liu}, \&
  {Cheung}}]{2022A&A...667A.166C}
{Chitta}, L.~P., {Peter}, H., {Parenti}, S., {et~al.} 2022,
  \bibinfo{title}{{Solar coronal heating from small-scale magnetic braids},}
  \aap, 667, A166, \dodoi{10.1051/0004-6361/202244170}

\bibitem[{L.~P. {Chitta} {et~al.}(2026){Chitta}, {Pontin}, {Priest},
  {Berghmans}, {Kraaikamp}, {Rodriguez}, {Verbeeck}, {Zhukov}, {Krucker},
  {Aznar Cuadrado}, {Calchetti}, {Hirzberger}, {Peter}, {Sch{\"u}hle},
  {Solanki}, {Teriaca}, {Giunta}, {Auch{\`e}re}, {Harra}, \&
  {M{\"u}ller}}]{2026A&A...705A.113C}
{Chitta}, L.~P., {Pontin}, D.~I., {Priest}, E.~R., {et~al.} 2026,
  \bibinfo{title}{{A magnetic avalanche as the central engine powering a solar
  flare},} \aap, 705, A113, \dodoi{10.1051/0004-6361/202557253}

\bibitem[{H. {Collier} {et~al.}(2024){Collier}, {Hayes}, {Purkhart}, {Krucker},
  {Ryan}, {Polito}, {Veronig}, {Harra}, {Berghmans}, {Kraaikamp}, {Dominique},
  {Dolla}, \& {Verbeeck}}]{2024A&A...692A.176C}
{Collier}, H., {Hayes}, L.~A., {Purkhart}, S., {et~al.} 2024,
  \bibinfo{title}{{Solar flares in the Solar Orbiter era: Short-exposure
  EUI/FSI observations of STIX flares},} \aap, 692, A176,
  \dodoi{10.1051/0004-6361/202451838}

\bibitem[{M.~F. {Corchado Albelo} {et~al.}(2024){Corchado Albelo},
  {Kazachenko}, \& {Lynch}}]{2024ApJ...965...16C}
{Corchado Albelo}, M.~F., {Kazachenko}, M.~D., \& {Lynch}, B.~J. 2024,
  \bibinfo{title}{{Inferring Fundamental Properties of the Flare Current Sheet
  Using Flare Ribbons: Oscillations in the Reconnection Flux Rates},} \apj,
  965, 16, \dodoi{10.3847/1538-4357/ad25f4}

\bibitem[{J.~T. {Dahlin} {et~al.}(2025){Dahlin}, {Antiochos}, {Wyper}, {Qiu},
  \& {DeVore}}]{2025ApJ...993...31D}
{Dahlin}, J.~T., {Antiochos}, S.~K., {Wyper}, P.~F., {Qiu}, J., \& {DeVore},
  C.~R. 2025, \bibinfo{title}{{Determining the 3D Dynamics of Solar Flare
  Magnetic Reconnection},} \apj, 993, 31, \dodoi{10.3847/1538-4357/ae03c5}

\bibitem[{J.~C. {del Toro Iniesta} {et~al.}(2025){del Toro Iniesta}, {Orozco
  Su{\'a}rez}, {{\'A}lvarez-Herrero}, {Sanchis Kilders}, {P{\'e}rez-Grande},
  {Ruiz Cobo}, {Bellot Rubio}, {Balaguer Jim{\'e}nez}, {L{\'o}pez Jim{\'e}nez},
  {{\'A}lvarez Garc{\'\i}a}, {Ramos M{\'a}s}, {Cobos Carrascosa}, {Labrousse},
  {Moreno Mantas}, {Morales-Fern{\'a}ndez}, {Aparicio del Moral}, {S{\'a}nchez
  G{\'o}mez}, {Bail{\'o}n Mart{\'\i}nez}, {Bail{\'e}n}, {Strecker},
  {Siu-Tapia}, {Santamarina Guerrero}, {Moreno Vacas}, {Ati{\'e}nzar
  Garc{\'\i}a}, {Dorantes Monteagudo}, {Bustamante}, {Tobaruela},
  {Fern{\'a}ndez-Medina}, {N{\'u}{\~n}ez Peral}, {Cebollero},
  {Garranzo-Garc{\'\i}a}, {Garc{\'\i}a Parejo}, {Gonzalo Melchor}, {S{\'a}nchez
  Rodr{\'\i}guez}, {Campos-Jara}, {Laguna}, {Silva-L{\'o}pez}, {Blanco
  Rodr{\'\i}guez}, {Gasent Blesa}, {Rodr{\'\i}guez Mart{\'\i}nez}, {Ferreres},
  {Gilabert Palmer}, {Torralbo}, {Piqueras}, {Gonz{\'a}lez-B{\'a}rcena},
  {Fern{\'a}ndez}, {Hern{\'a}ndez Exp{\'o}sito}, {P{\'a}ez Ma{\~n}{\'a}},
  {Magdaleno Castell{\'o}}, {Rodr{\'\i}guez Valido}, {Korpi-Lagg}, {Gandorfer},
  {Solanki}, {Berkefeld}, {Bernasconi}, {Feller}, {Katsukawa},
  {Riethm{\"u}ller}, {Smitha}, {Kubo}, {Mart{\'\i}nez Pillet}, {Grauf}, {Bell},
  \& {Carpenter}}]{2025SoPh..300..148D}
{del Toro Iniesta}, J.~C., {Orozco Su{\'a}rez}, D., {{\'A}lvarez-Herrero}, A.,
  {et~al.} 2025, \bibinfo{title}{{TuMag: The Tunable Magnetograph for the
  SUNRISE III Mission},} \solphys, 300, 148, \dodoi{10.1007/s11207-025-02562-5}

\bibitem[{M. {Druett} {et~al.}(2024){Druett}, {Ruan}, \&
  {Keppens}}]{2024A&A...684A.171D}
{Druett}, M., {Ruan}, W., \& {Keppens}, R. 2024, \bibinfo{title}{{Exploring
  self-consistent 2.5D flare simulations with MPI-AMRVAC},} \aap, 684, A171,
  \dodoi{10.1051/0004-6361/202347600}

\bibitem[{J. {Dud{\'\i}k} {et~al.}(2016){Dud{\'\i}k}, {Polito}, {Janvier},
  {Mulay}, {Karlick{\'y}}, {Aulanier}, {Del Zanna}, {Dzif{\v{c}}{\'a}kov{\'a}},
  {Mason}, \& {Schmieder}}]{2016ApJ...823...41D}
{Dud{\'\i}k}, J., {Polito}, V., {Janvier}, M., {et~al.} 2016,
  \bibinfo{title}{{Slipping Magnetic Reconnection, Chromospheric Evaporation,
  Implosion, and Precursors in the 2014 September 10 X1.6-Class Solar Flare},}
  \apj, 823, 41, \dodoi{10.3847/0004-637X/823/1/41}

\bibitem[{L. {Fletcher} {et~al.}(2013){Fletcher}, {Hannah}, {Hudson}, \&
  {Innes}}]{2013ApJ...771..104F}
{Fletcher}, L., {Hannah}, I.~G., {Hudson}, H.~S., \& {Innes}, D.~E. 2013,
  \bibinfo{title}{{Flare Ribbon Energetics in the Early Phase of an SDO
  Flare},} \apj, 771, 104, \dodoi{10.1088/0004-637X/771/2/104}

\bibitem[{S.~L. {Freeland} \& B.~N. {Handy}(1998){Freeland} \&
  {Handy}}]{1998SoPh..182..497F}
{Freeland}, S.~L., \& {Handy}, B.~N. 1998, \bibinfo{title}{{Data Analysis with
  the SolarSoft System},} \solphys, 182, 497, \dodoi{10.1023/A:1005038224881}

\bibitem[{R.~J. {French} {et~al.}(2025){French}, {Kazachenko}, {Berghmans},
  {D'Huys}, {Dominique}, {Patel}, {Talpeanu}, {Tamburri}, \&
  {Yadav}}]{2025ApJ...995L..54F}
{French}, R.~J., {Kazachenko}, M.~D., {Berghmans}, D., {et~al.} 2025,
  \bibinfo{title}{{Evolution of Flare Ribbon Bead-like Structures in a Solar
  Flare},} \apjl, 995, L54, \dodoi{10.3847/2041-8213/ae2684}

\bibitem[{H. {Hu} {et~al.}(2022){Hu}, {Liu}, {Chitta}, {Peter}, \&
  {Ding}}]{2022ApJ...940L..12H}
{Hu}, H., {Liu}, Y.~D., {Chitta}, L.~P., {Peter}, H., \& {Ding}, M. 2022,
  \bibinfo{title}{{Spectroscopic and Imaging Observations of Spatially Extended
  Magnetic Reconnection in the Splitting of a Solar Filament Structure},}
  \apjl, 940, L12, \dodoi{10.3847/2041-8213/ac9dfd}

\bibitem[{J. {Jing} {et~al.}(2016){Jing}, {Xu}, {Cao}, {Liu}, {Gary}, \&
  {Wang}}]{2016NatSR...624319J}
{Jing}, J., {Xu}, Y., {Cao}, W., {et~al.} 2016, \bibinfo{title}{{Unprecedented
  Fine Structure of a Solar Flare Revealed by the 1.6{\,}m New Solar
  Telescope},} Scientific Reports, 6, 24319, \dodoi{10.1038/srep24319}

\bibitem[{M.~D. {Kazachenko}(2023){Kazachenko}}]{2023ApJ...958..104K}
{Kazachenko}, M.~D. 2023, \bibinfo{title}{{A Database of Magnetic and
  Thermodynamic Properties of Confined and Eruptive Solar Flares},} \apj, 958,
  104, \dodoi{10.3847/1538-4357/ad004e}

\bibitem[{M.~D. {Kazachenko} {et~al.}(2017){Kazachenko}, {Lynch}, {Welsch}, \&
  {Sun}}]{2017ApJ...845...49K}
{Kazachenko}, M.~D., {Lynch}, B.~J., {Welsch}, B.~T., \& {Sun}, X. 2017,
  \bibinfo{title}{{A Database of Flare Ribbon Properties from the Solar
  Dynamics Observatory. I. Reconnection Flux},} \apj, 845, 49,
  \dodoi{10.3847/1538-4357/aa7ed6}

\bibitem[{A. {Korpi-Lagg} {et~al.}(2025){Korpi-Lagg}, {Gandorfer}, {Solanki},
  {del Toro Iniesta}, {Katsukawa}, {Bernasconi}, {Berkefeld}, {Feller},
  {Riethm{\"u}ller}, {{\'A}lvarez-Herrero}, {Kubo}, {Mart{\'\i}nez Pillet},
  {Smitha}, {Orozco Su{\'a}rez}, {Grauf}, {Carpenter}, {Bell},
  {{\'A}lvarez-Alonso}, {{\'A}lvarez Garc{\'\i}a}, {Aparicio del Moral},
  {Ati{\'e}nzar}, {Ayoub}, {Bail{\'e}n}, {Bail{\'o}n Mart{\'\i}nez}, {Balaguer
  Jim{\'e}nez}, {Barthol}, {Bayon Laguna}, {Bellot Rubio}, {Bergmann}, {Blanco
  Rodr{\'\i}guez}, {Bochmann}, {Borrero}, {Campos-Jara}, {Castellanos
  Dur{\'a}n}, {Cebollero}, {Conde Rodr{\'\i}guez}, {Deutsch}, {Eaton},
  {Fern{\'a}ndez-Medina}, {Fernandez-Rico}, {Ferreres}, {Garc{\'\i}a},
  {Garc{\'\i}a Alarcia}, {Garc{\'\i}a Parejo}, {Garranzo-Garc{\'\i}a}, {Gasent
  Blesa}, {Gerber}, {Germerott}, {Gilabert Palmer}, {Gizon}, {G{\'o}mez
  S{\'a}nchez-Tirado}, {Gonz{\'a}lez-B{\'a}rcena}, {Gonzalo Melchor},
  {Goodyear}, {Hara}, {Harnes}, {Heerlein}, {Heidecke}, {Heinrichs},
  {Hern{\'a}ndez Exp{\'o}sito}, {Hirzberger}, {Hoelken}, {Hyun}, {Iglesias},
  {Ishikawa}, {Jeon}, {Kawabata}, {Kolleck}, {Laguna}, {Lomas}, {L{\'o}pez
  Jim{\'e}nez}, {Manzano}, {Matsumoto}, {Mayo Turrado}, {Meierdierks},
  {Meining}, {Monecke}, {Morales-Fern{\'a}ndez}, {Moreno Mantas}, {Moreno
  Vacas}, {M{\"u}ller}, {M{\"u}ller}, {Naito}, {Nakai}, {N{\'u}{\~n}ez Peral},
  {Oba}, {Palo}, {P{\'e}rez-Grande}, {Piqueras Carre{\~n}o}, {Preis},
  {Przybylski}, {Quintero Noda}, {Ramanath}, {Ramos M{\'a}s}, {Raouafi},
  {Rivas-Mart{\'\i}nez}, {Rodr{\'\i}guez Mart{\'\i}nez}, {Rodr{\'\i}guez
  Valido}, {Ruiz Cobo}, {S{\'a}nchez Rodr{\'\i}guez}, {Sanchez Toledo},
  {S{\'a}nchez G{\'o}mez}, {Sanchis Kilders}, {Sant}, {Santamarina Guerrero},
  {Schulze}, {Shimizu}, {Silva-L{\'o}pez}, {Singh}, {Siu-Tapia}, {Sonner},
  {Staub}, {Strecker}, {Tobaruela}, {Torralbo}, {Tritschler}, {Tsuzuki},
  {Uraguchi}, {Volkmer}, {Vourlidas}, {Vukadinovi{\'c}}, {Werner}, \&
  {Zerr}}]{2025SoPh..300...75K}
{Korpi-Lagg}, A., {Gandorfer}, A., {Solanki}, S.~K., {et~al.} 2025,
  \bibinfo{title}{{SUNRISE III: Overview of Observatory and Instruments},}
  \solphys, 300, 75, \dodoi{10.1007/s11207-025-02485-1}

\bibitem[{C. {Kuckein} {et~al.}(2025){Kuckein}, {Collados}, {Asensio Ramos},
  {D{\'\i}az Baso}, {Felipe}, {Quintero Noda}, {Kleint}, {Fletcher}, \&
  {Matthews}}]{2025A&A...699A.121K}
{Kuckein}, C., {Collados}, M., {Asensio Ramos}, A., {et~al.} 2025,
  \bibinfo{title}{{Inferring chromospheric velocities in an M3.2 flare using He
  I 1083.0 nm and Ca II 854.2 nm},} \aap, 699, A121,
  \dodoi{10.1051/0004-6361/202554798}

\bibitem[{J.~R. {Lemen} {et~al.}(2012){Lemen}, {Title}, {Akin}, {Boerner},
  {Chou}, {Drake}, {Duncan}, {Edwards}, {Friedlaender}, {Heyman}, {Hurlburt},
  {Katz}, {Kushner}, {Levay}, {Lindgren}, {Mathur}, {McFeaters}, {Mitchell},
  {Rehse}, {Schrijver}, {Springer}, {Stern}, {Tarbell}, {Wuelser}, {Wolfson},
  {Yanari}, {Bookbinder}, {Cheimets}, {Caldwell}, {Deluca}, {Gates}, {Golub},
  {Park}, {Podgorski}, {Bush}, {Scherrer}, {Gummin}, {Smith}, {Auker},
  {Jerram}, {Pool}, {Soufli}, {Windt}, {Beardsley}, {Clapp}, {Lang}, \&
  {Waltham}}]{2012SoPh..275...17L}
{Lemen}, J.~R., {Title}, A.~M., {Akin}, D.~J., {et~al.} 2012,
  \bibinfo{title}{{The Atmospheric Imaging Assembly (AIA) on the Solar Dynamics
  Observatory (SDO)},} \solphys, 275, 17, \dodoi{10.1007/s11207-011-9776-8}

\bibitem[{C. {Li} {et~al.}(2022){Li}, {Fang}, {Li}, {Ding}, {Chen}, {Qiu},
  {You}, {Yuan}, {An}, {Tao}, {Li}, {Chen}, {Liu}, {Mei}, {Yang}, {Zhang},
  {Cheng}, {Chen}, {Chen}, {Gu}, {Huang}, {Liu}, {Han}, {Xin}, {Chen}, {Ni},
  {Wang}, {Rao}, {Li}, {Lu}, {Wang}, {Lin}, {Jiang}, {Meng}, \&
  {Zhao}}]{2022SCPMA..6589602L}
{Li}, C., {Fang}, C., {Li}, Z., {et~al.} 2022, \bibinfo{title}{{The Chinese
  H{\ensuremath{\alpha}} Solar Explorer (CHASE) mission: An overview},} Science
  China Physics, Mechanics, and Astronomy, 65, 289602,
  \dodoi{10.1007/s11433-022-1893-3}

\bibitem[{J. {L{\"o}rin{\v{c}}{\'\i}k} {et~al.}(2025){L{\"o}rin{\v{c}}{\'\i}k},
  {Polito}, {Kerr}, {Hayes}, \& {Russell}}]{2025ApJ...986...73L}
{L{\"o}rin{\v{c}}{\'\i}k}, J., {Polito}, V., {Kerr}, G.~S., {Hayes}, L.~A., \&
  {Russell}, A. J.~B. 2025, \bibinfo{title}{{Probing Progression of Heating
  Through the Lower Flare Atmosphere via High-cadence IRIS Spectroscopy},}
  \apj, 986, 73, \dodoi{10.3847/1538-4357/adccc8}

\bibitem[{V. {Mart{\'\i}nez Pillet}(2000){Mart{\'\i}nez
  Pillet}}]{2000A&A...361..734M}
{Mart{\'\i}nez Pillet}, V. 2000, \bibinfo{title}{{Spectral signature of
  uncombed penumbral magnetic fields},} \aap, 361, 734

\bibitem[{P.~J. {Mauas} {et~al.}(1988){Mauas}, {Avrett}, \&
  {Loeser}}]{1988ApJ...330.1008M}
{Mauas}, P.~J., {Avrett}, E.~H., \& {Loeser}, R. 1988, \bibinfo{title}{{MG i as
  a Probe of the Solar Chromosphere: The Atomic Model},} \apj, 330, 1008,
  \dodoi{10.1086/166530}

\bibitem[{A.~J. {Monson} {et~al.}(2024){Monson}, {Mathioudakis}, \&
  {Kowalski}}]{2024ApJ...963...40M}
{Monson}, A.~J., {Mathioudakis}, M., \& {Kowalski}, A.~F. 2024,
  \bibinfo{title}{{Deconstructing Photospheric Spectral Lines in Solar and
  Stellar Flares},} \apj, 963, 40, \dodoi{10.3847/1538-4357/ad16da}

\bibitem[{R.~L. {Moore} {et~al.}(2001){Moore}, {Sterling}, {Hudson}, \&
  {Lemen}}]{2001ApJ...552..833M}
{Moore}, R.~L., {Sterling}, A.~C., {Hudson}, H.~S., \& {Lemen}, J.~R. 2001,
  \bibinfo{title}{{Onset of the Magnetic Explosion in Solar Flares and Coronal
  Mass Ejections},} \apj, 552, 833, \dodoi{10.1086/320559}

\bibitem[{H. {Morgan} \& M. {Druckm{\"u}ller}(2014){Morgan} \&
  {Druckm{\"u}ller}}]{2014SoPh..289.2945M}
{Morgan}, H., \& {Druckm{\"u}ller}, M. 2014, \bibinfo{title}{{Multi-Scale
  Gaussian Normalization for Solar Image Processing},} \solphys, 289, 2945,
  \dodoi{10.1007/s11207-014-0523-9}

\bibitem[{B. {Panos} {et~al.}(2018){Panos}, {Kleint}, {Huwyler}, {Krucker},
  {Melchior}, {Ullmann}, \& {Voloshynovskiy}}]{2018ApJ...861...62P}
{Panos}, B., {Kleint}, L., {Huwyler}, C., {et~al.} 2018,
  \bibinfo{title}{{Identifying Typical Mg II Flare Spectra Using Machine
  Learning},} \apj, 861, 62, \dodoi{10.3847/1538-4357/aac779}

\bibitem[{W.~D. {Pesnell} {et~al.}(2012){Pesnell}, {Thompson}, \&
  {Chamberlin}}]{2012SoPh..275....3P}
{Pesnell}, W.~D., {Thompson}, B.~J., \& {Chamberlin}, P.~C. 2012,
  \bibinfo{title}{{The Solar Dynamics Observatory (SDO)},} \solphys, 275, 3,
  \dodoi{10.1007/s11207-011-9841-3}

\bibitem[{A.~G.~M. {Pietrow} {et~al.}(2022){Pietrow}, {Druett}, {de la Cruz
  Rodriguez}, {Calvo}, \& {Kiselman}}]{2022A&A...659A..58P}
{Pietrow}, A.~G.~M., {Druett}, M.~K., {de la Cruz Rodriguez}, J., {Calvo}, F.,
  \& {Kiselman}, D. 2022, \bibinfo{title}{{Physical properties of a fan-shaped
  jet backlit by an X9.3 flare},} \aap, 659, A58,
  \dodoi{10.1051/0004-6361/202142346}

\bibitem[{A.~G.~M. {Pietrow} {et~al.}(2024){Pietrow}, {Druett}, \&
  {Singh}}]{2024A&A...685A.137P}
{Pietrow}, A.~G.~M., {Druett}, M.~K., \& {Singh}, V. 2024,
  \bibinfo{title}{{Spectral variations within solar flare ribbons},} \aap, 685,
  A137, \dodoi{10.1051/0004-6361/202348839}

\bibitem[{V. {Polito} {et~al.}(2023){Polito}, {Kerr}, {Xu}, {Sadykov}, \&
  {Lorincik}}]{2023ApJ...944..104P}
{Polito}, V., {Kerr}, G.~S., {Xu}, Y., {Sadykov}, V.~M., \& {Lorincik}, J.
  2023, \bibinfo{title}{{Solar Flare Ribbon Fronts. I. Constraining Flare
  Energy Deposition with IRIS Spectroscopy},} \apj, 944, 104,
  \dodoi{10.3847/1538-4357/acaf7c}

\bibitem[{E.~R. {Priest} \& T.~G. {Forbes}(2002){Priest} \&
  {Forbes}}]{2002A&ARv..10..313P}
{Priest}, E.~R., \& {Forbes}, T.~G. 2002, \bibinfo{title}{{The magnetic nature
  of solar flares},} \aapr, 10, 313, \dodoi{10.1007/s001590100013}

\bibitem[{J. {Qiu} \& J. {Cheng}(2022){Qiu} \& {Cheng}}]{2022SoPh..297...80Q}
{Qiu}, J., \& {Cheng}, J. 2022, \bibinfo{title}{{Properties and Energetics of
  Magnetic Reconnection: I. Evolution of Flare Ribbons},} \solphys, 297, 80,
  \dodoi{10.1007/s11207-022-02003-7}

\bibitem[{Y. {Qiu} {et~al.}(2022){Qiu}, {Rao}, {Li}, {Fang}, {Ding}, {Li},
  {Ni}, {Wang}, {Hong}, {Hao}, {Dai}, {Chen}, {Wan}, {Xu}, {You}, {Yuan},
  {Tao}, {Li}, {He}, \& {Liu}}]{2022SCPMA..6589603Q}
{Qiu}, Y., {Rao}, S., {Li}, C., {et~al.} 2022, \bibinfo{title}{{Calibration
  procedures for the CHASE/HIS science data},} Science China Physics,
  Mechanics, and Astronomy, 65, 289603, \dodoi{10.1007/s11433-022-1900-5}

\bibitem[{K.~P. {Reardon} {et~al.}(2011){Reardon}, {Wang}, {Muglach}, \&
  {Warren}}]{2011ApJ...742..119R}
{Reardon}, K.~P., {Wang}, Y.-M., {Muglach}, K., \& {Warren}, H.~P. 2011,
  \bibinfo{title}{{Evidence for Two Separate but Interlaced Components of the
  Chromospheric Magnetic Field},} \apj, 742, 119,
  \dodoi{10.1088/0004-637X/742/2/119}

\bibitem[{W. {Ruan} {et~al.}(2023){Ruan}, {Yan}, \&
  {Keppens}}]{2023ApJ...947...67R}
{Ruan}, W., {Yan}, L., \& {Keppens}, R. 2023,
  \bibinfo{title}{{Magnetohydrodynamic Turbulence Formation in Solar Flares: 3D
  Simulation and Synthetic Observations},} \apj, 947, 67,
  \dodoi{10.3847/1538-4357/ac9b4e}

\bibitem[{F. {Rubio da Costa} \& L. {Kleint}(2017){Rubio da Costa} \&
  {Kleint}}]{2017ApJ...842...82R}
{Rubio da Costa}, F., \& {Kleint}, L. 2017, \bibinfo{title}{{A Parameter Study
  for Modeling Mg II h and k Emission during Solar Flares},} \apj, 842, 82,
  \dodoi{10.3847/1538-4357/aa6eaf}

\bibitem[{R.~J. {Rutten} {et~al.}(2011){Rutten}, {Leenaarts}, {Rouppe van der
  Voort}, {de Wijn}, {Carlsson}, \& {Hansteen}}]{2011A&A...531A..17R}
{Rutten}, R.~J., {Leenaarts}, J., {Rouppe van der Voort}, L.~H.~M., {et~al.}
  2011, \bibinfo{title}{{Quiet-Sun imaging asymmetries in Na I D$_{1}$ compared
  with other strong Fraunhofer lines},} \aap, 531, A17,
  \dodoi{10.1051/0004-6361/201116984}

\bibitem[{D.~F. {Ryan} {et~al.}(2025){Ryan}, {Hayes}, {Collier}, {Kerr},
  {Inglis}, {Williams}, {Walsh}, {Janvier}, {M{\"u}ller}, {Berghmans},
  {Verbeeck}, {Kraaikamp}, {Young}, {Kucera}, {Krucker}, {Stiefel},
  {Calchetti}, {Reeves}, {Savage}, \& {Polito}}]{2025SoPh..300..152R}
{Ryan}, D.~F., {Hayes}, L.~A., {Collier}, H., {et~al.} 2025,
  \bibinfo{title}{{Solar Orbiter's 2024 Major Flare Campaigns: An Overview},}
  \solphys, 300, 152, \dodoi{10.1007/s11207-025-02561-6}

\bibitem[{P.~H. {Scherrer} {et~al.}(2012){Scherrer}, {Schou}, {Bush},
  {Kosovichev}, {Bogart}, {Hoeksema}, {Liu}, {Duvall}, {Zhao}, {Title},
  {Schrijver}, {Tarbell}, \& {Tomczyk}}]{2012SoPh..275..207S}
{Scherrer}, P.~H., {Schou}, J., {Bush}, R.~I., {et~al.} 2012,
  \bibinfo{title}{{The Helioseismic and Magnetic Imager (HMI) Investigation for
  the Solar Dynamics Observatory (SDO)},} \solphys, 275, 207,
  \dodoi{10.1007/s11207-011-9834-2}

\bibitem[{K. {Shibata} \& T. {Magara}(2011){Shibata} \&
  {Magara}}]{2011LRSP....8....6S}
{Shibata}, K., \& {Magara}, T. 2011, \bibinfo{title}{{Solar Flares:
  Magnetohydrodynamic Processes},} Living Reviews in Solar Physics, 8, 6,
  \dodoi{10.12942/lrsp-2011-6}

\bibitem[{A.~L. {Siu-Tapia} {et~al.}(2025{\natexlab{a}}){Siu-Tapia}, {Bellot
  Rubio}, \& {Orozco Su{\'a}rez}}]{2025A&A...696A.105S}
{Siu-Tapia}, A.~L., {Bellot Rubio}, L.~R., \& {Orozco Su{\'a}rez}, D.
  2025{\natexlab{a}}, \bibinfo{title}{{Diagnosing the solar atmosphere through
  the Mg I b$_{2}$ 5173 {\r{A}} line: I. Nonlocal thermodynamic equilibrium
  inversions versus traditional inferences},} \aap, 696, A105,
  \dodoi{10.1051/0004-6361/202453230}

\bibitem[{A.~L. {Siu-Tapia} {et~al.}(2025{\natexlab{b}}){Siu-Tapia}, {Bellot
  Rubio}, {Orozco Su{\'a}rez}, \& {Gafeira}}]{2025A&A...696A.106S}
{Siu-Tapia}, A.~L., {Bellot Rubio}, L.~R., {Orozco Su{\'a}rez}, D., \&
  {Gafeira}, R. 2025{\natexlab{b}}, \bibinfo{title}{{Diagnosing the solar
  atmosphere through the Mg I b$_{2}$ 5173 {\r{A}} line: II. Morphological
  classification of the intensity and circular polarization profiles},} \aap,
  696, A106, \dodoi{10.1051/0004-6361/202453232}

\bibitem[{S.~K. {Solanki} \& C.~A.~P. {Montavon}(1993){Solanki} \&
  {Montavon}}]{1993A&A...275..283S}
{Solanki}, S.~K., \& {Montavon}, C.~A.~P. 1993, \bibinfo{title}{{Uncombed
  fields as the source of the broad-band circular polarization of sunspots.},}
  \aap, 275, 283

\bibitem[{S.~K. {Solanki} {et~al.}(2010){Solanki}, {Barthol}, {Danilovic},
  {Feller}, {Gandorfer}, {Hirzberger}, {Riethm{\"u}ller}, {Sch{\"u}ssler},
  {Bonet}, {Mart{\'\i}nez Pillet}, {del Toro Iniesta}, {Domingo}, {Palacios},
  {Kn{\"o}lker}, {Bello Gonz{\'a}lez}, {Berkefeld}, {Franz}, {Schmidt}, \&
  {Title}}]{2010ApJ...723L.127S}
{Solanki}, S.~K., {Barthol}, P., {Danilovic}, S., {et~al.} 2010,
  \bibinfo{title}{{SUNRISE: Instrument, Mission, Data, and First Results},}
  \apjl, 723, L127, \dodoi{10.1088/2041-8205/723/2/L127}

\bibitem[{S.~K. {Solanki} {et~al.}(2017){Solanki}, {Riethm{\"u}ller},
  {Barthol}, {Danilovic}, {Deutsch}, {Doerr}, {Feller}, {Gandorfer},
  {Germerott}, {Gizon}, {Grauf}, {Heerlein}, {Hirzberger}, {Kolleck}, {Lagg},
  {Meller}, {Tomasch}, {van Noort}, {Blanco Rodr{\'\i}guez}, {Gasent Blesa},
  {Balaguer Jim{\'e}nez}, {Del Toro Iniesta}, {L{\'o}pez Jim{\'e}nez}, {Orozco
  Suarez}, {Berkefeld}, {Halbgewachs}, {Schmidt}, {{\'A}lvarez-Herrero},
  {Sabau-Graziati}, {P{\'e}rez Grande}, {Mart{\'\i}nez Pillet}, {Card},
  {Centeno}, {Kn{\"o}lker}, \& {Lecinski}}]{2017ApJS..229....2S}
{Solanki}, S.~K., {Riethm{\"u}ller}, T.~L., {Barthol}, P., {et~al.} 2017,
  \bibinfo{title}{{The Second Flight of the Sunrise Balloon-borne Solar
  Observatory: Overview of Instrument Updates, the Flight, the Data, and First
  Results},} \apjs, 229, 2, \dodoi{10.3847/1538-4365/229/1/2}

\bibitem[{S.~K. Solanki {et~al.}(2026)Solanki, Smitha, Lagg, Gandorfer, del
  Toro~Iniesta, Katsukawa, Bernasconi, Berkefeld, Feller, Riethmüller,
  Álvarez Herrero, Kubo, Suárez, Grauf, Carpenter, Bell, Pillet, Gizon,
  Bailén, Rodríguez, Durán, Harnes, Hoelken, Iglesias, Ishikawa, Kawabata,
  Matsumoto, Oba, Singh, Siu-Tapia, Strecker, Vukadinović, van Noort,
  Jiménez, Kilders, Torralbo, Kuckein, Hara, Shimizu, Volkmer, Preis, Raouafi,
  Vourlidas, Hirzberger, Deutsch, Germerott, Heerlein, Kolleck, Álvarez
  García, Jiménez, Rubio, Morales-Fernández, Mantas, del Moral, Gómez,
  Martínez, Guerrero, Expósito, Tobaruela, Blesa, Schulze, Eaton, Palo,
  Ayoub, Naito, Noda, Uraguchi, Tsuzuki, \& Carreño}]{solankietal2026}
Solanki, S.~K., Smitha, H.~N., Lagg, A., {et~al.} 2026, Sunrise III:
  Instrument, mission, data, and first results, \doarXiv{2606.07989}

\bibitem[{C.~A. {Tamburri} {et~al.}(2025){Tamburri}, {Kazachenko}, {Cauzzi},
  {Kowalski}, {French}, {Yadav}, {Evans}, {Notsu}, {Corchado-Albelo},
  {Reardon}, \& {Tritschler}}]{2025ApJ...990L...3T}
{Tamburri}, C.~A., {Kazachenko}, M.~D., {Cauzzi}, G., {et~al.} 2025,
  \bibinfo{title}{{Unveiling Unprecedented Fine Structure in Coronal Flare
  Loops with the DKIST},} \apjl, 990, L3, \dodoi{10.3847/2041-8213/adf95e}

\bibitem[{J. {Thoen Faber} {et~al.}(2025){Thoen Faber}, {Joshi}, {Rouppe van
  der Voort}, {Wedemeyer}, {Fletcher}, {Aulanier}, \&
  {N{\'o}brega-Siverio}}]{2025A&A...693A...8T}
{Thoen Faber}, J., {Joshi}, R., {Rouppe van der Voort}, L., {et~al.} 2025,
  \bibinfo{title}{{High-resolution observational analysis of flare ribbon fine
  structures},} \aap, 693, A8, \dodoi{10.1051/0004-6361/202452370}

\bibitem[{P.~F. {Wyper} \& D.~I. {Pontin}(2021){Wyper} \&
  {Pontin}}]{2021ApJ...920..102W}
{Wyper}, P.~F., \& {Pontin}, D.~I. 2021, \bibinfo{title}{{Is Flare Ribbon Fine
  Structure Related to Tearing in the Flare Current Sheet?},} \apj, 920, 102,
  \dodoi{10.3847/1538-4357/ac1943}

\bibitem[{P.~R. {Young} {et~al.}(2013){Young}, {Doschek}, {Warren}, \&
  {Hara}}]{2013ApJ...766..127Y}
{Young}, P.~R., {Doschek}, G.~A., {Warren}, H.~P., \& {Hara}, H. 2013,
  \bibinfo{title}{{Properties of a Solar Flare Kernel Observed by Hinode and
  SDO},} \apj, 766, 127, \dodoi{10.1088/0004-637X/766/2/127}

\end{thebibliography}

\end{document}